\documentclass[12pt ,openright]{book}
\def\Ai{\;{\rm Ai}\;}
\def\sgn{\;{\rm sgn}\;}
\def\implies{\Rightarrow}

\def\ignore#1{\ }
\def\hattilde#1{\hat{\tilde #1}}
\newcommand{\m}[1]{ { $#1$} }

\newcommand{\beq}{ \begin{eqnarray}}
\newcommand{\eeq}{\end{eqnarray}}
\newcommand{\numeq}{\end{eqnarray}}

\newcommand{\half}{ {1\over 2} }

\newcommand{\pdr}{\partial}
\newcommand{\beqs}{\begin{eqnarray}}
\newcommand{\eeqs}{\nonumber\end{eqnarray}}

\newcommand{\eps}{\epsilon}

\usepackage{graphics}

\def\tr{\;{\rm tr}}

\def\Gr{\;{\rm Gr}}

\newcounter{example}

\newcounter{remark}

\begin{document}
\frontmatter
\begin{center}
{\Huge Derivation of the  Hadronic Structure Functions }
{\Huge From}
{\Huge Quantum Chromodynamics}

{Lectures at the the Feza G\"ursey Institute, Istanbul\\}
 {and the Mittag-Leffler Institute of the Royal Swedish Academy 
of Sciences, Stockholm in 1998.}
\end{center}
\centerline{\Large
S. G. Rajeev\footnote{rajeev@pas.rochester.edu}}

\begin{center}
   {\it Department of Physics and Astronomy, University of Rochester, 
    Rochester, New York 14627} \\
\end{center}

\centerline{\bf\Large   Abstract}

\vspace{0.5cm}

We solve a long-standing problem in particle physics: that of deriving the
Deep Inelastic  structure functions of the proton from the fundamental theory
of strong interactions, Quantum ChromoDynamics (QCD). In the Bjorken limit, 
the momenta of the constituents of the proton (the partons) 
can be assumed to be in  a two-dimensional plane in Minkowski space:
 a dimensional reduction of QCD to two space-time 
dimensions. Two dimensional QCD is then shown to be equivalent for all
energies and values of number of colors \m{N} to  a new theory of
hadrons,  Quantum HadronDynamics (QHD). The phase space of QHD is
 the  Grassmannian (set of subspcaes) of
the complex Hilbert space \m{L^2(R)}. The natural symplectic form
along with a hamiltonian define a classical dynamical system, which is
equivalent to the large \m{N} limit of QCD. 't Hooft's planar limit
 is
the linear approximation to our theory: we recover his integral
equation for the meson spectrum but also all the interactions of the mesons.
The Grassmannian is a union of
connected components labelled by an integer (the renormalized
dimension of the subspace) which has the physical meaning of baryon
number. The proton is the  topological soliton: 
the minimum of the energy in the sector with baryon number one 
gives the structure functions of the proton. We solve the resulting
integral equations numerically; the 
agreement with experimental data 
is quite good for values of  the Bjorken variable \m{x_B>0.2}.

\frontmatter

\tableofcontents
\listoffigures

\mainmatter

\chapter{Introduction}

The first indication that the atom contains a point--like nucleus came
 from experiments of Rutherford. He scattered alpha particles (which
 are positively charged) from a gold foil. Since the negative charges
 (electrons) were known to be much lighter than the alpha
 particles,they would not scatter the alpha particles very much. If
 the positive charges inside the atom were more or less uniformly
 distributed, the alpha particles would not be scattered through wide
 angles.  Rutherford found to the contrary that the alpha particles
 were scattered by wide angles. In fact the probability of scattering
 through an angle \m{\theta} is proportional to
 \m{\sin^{-4}{\theta\over 2}}, exactly what would happen if all the
 positive charge of the atom were concentrated at a point: he had
 discovered the atomic nucleus.  Soon after, it was realized that the
 nucleus is of finite size, although small compared to the atom. It is
 composed of protons and neutrons bound together by the strong
 interaction.

Many years later, another celebrated series of experiments
 studied the structure
of the  proton itself. It was found that it was not an elementary
 particle either, in fact that it was made of point-like constituents. 
In such a  `Deep Inelastic Scattering' experiment
an  electron (or neutrino) beam is scattered  by a  proton (or
a nucleus containing both protons and neutrons). The idea is to learn about 
the still mysterious strong interactions  using the electroweak
 interactions
 as a probe.

A  particle that can take part in the strong
interactions is called a `hadron'. There are two kinds of hadrons: those of half-integer angular
momentum are called `baryons' and those of integer angular momentum
are the `mesons'. The proton is the lightest baryon and the \m{\pi}-meson 
(pion) the lightest
 meson. There are an infinite number of baryons and mesons but the more 
energetic ones are   unstable against decay into the lighter hadrons.
(We consider an idealized world in which the electromagnetic, weak and 
gravitational interactions of the hadrons with each other are ignored. Thus 
a particle is considered stable if its decays are purely electromagnetic or 
weak. This is the sense in which the pion is stable. )

The hadrons are composed of more elementary constituents: the
 quarks, anti-quarks  and gluons. These constituents are collectively
 called partons\cite{bjorken,feynman}.
However, it has been found experimentaly that it is impossible to create the 
partons in isolation: they only exist inside hadrons, and
thus, are 
not  true  particles. This is the phenomenon of `confinement'. More precisely,
only combinations of quarks and gluons invariant under the action of the color
 group \m{SU(N)} (see below) exist as isolation. Thus hadrons can be defined
 to be states that are invariant under color.  
Some authors use the word hadron to include 
 the quarks and gluons. We will use the term 
 `hadron'  to refer to   a   bound state of  quarks and gluons  which can exist
as an isolated particle. Of course such a particle may be unstable against
 decay into other hadrons.

Around the same time of these developments an entirely different
picture of a baryon was proposed by Skyrme \cite{skyrme}: that it is a 
topological
soliton made of an infinite number of mesons. Unfortunately, this idea
did not fit with the dominant view of the time and was largely
ignored. In the mid-eighties Skyrme's idea was finally revived by a
group at Syracuse University (including the author)
\cite{syracuseskyrme, rajeevthesis, baltasi} 
and integrated into the modern theory of strong
interactions, i.e., Quantum Chromodynamics. Witten \cite{witten} 
 clarified why  the baryons are
 fermions (when \m{N} is odd) even though the underlying theory has
 only bosonic fields. ( Witten had already 
arrived at the idea that baryons are solitons
in the large \m{N} limit of 't  Hooft \cite{thooftlargeN} by independent arguments 
\cite{wittenlargeN}. )

 But then the problem
remains how to reconcile it with the picture of a baryon as a bound
state of point-like constituents. We will show in these lectures how
the parton model can be derived from the soliton model. It fact it had
never been possible to derive the distribution functions of the
partons inside the hadron from fundamental principles. Our picture
will solve this problem, providing for the first time a quantitative
theory of the structure of a proton.

 The basic idea of our approach is to  find a new description of
 strong interactions directly in terms of hadrons rather than in terms
 of quarks and gluons. It should be equivalent to the color singlet
 sector of QCD at all energies and all values of the number of colors. 
We will call this theory quantum
 hadrondynamics. In four dimensions this is still just an idea; we
 havent been  able to show yet that such a new paradigm for strong
 interactions exists. However in the two dimensional case, I 
 constructed such an equivalent alternative formalism some years
 ago. Part of the motivation for studying the two dimensional case was
 that it would provide a solution to the problem of deriving the
 structure functions of  four dimensional QCD. (It has been known since
 the early work of Feynman and Bjorken that Deep Inelastic Scattering
 can be explained by a two dimensional theory of strong interactions.)
The first report of the idea of such an equivalent theory was in Ref. 
\cite{rajeevictp}. The solutions of the integral equations for the
 baryon wavefunction was first studied in
 Ref. \cite{rajeevmodphyslett}. A more detailed description of the
 theory appeared in Ref. \cite{2dhadron}. In another direction,
 these ideas were applied to spherically symmetric situations in
 Ref. \cite{sphqcd}. The methods of geometrical quantization were
 applied  to the problem in Ref. \cite{rajeevturgut}. There a version
 with bosonic quarks was also studied. We returned to the study of the
 baryon wavefunctions in Ref. \cite{krishraj} where the derivation of
 the parton model, already mentioned in \cite{2dhadron} was given in
 more detail.  In papers which will
 soon appear \cite{johnkrishraj}
  we will extend these results to include the anti-quark
 and sea quark distributions. Some of the ideas that go into this work
 are outlined in the last chapter. The basic mathematical ideas necessary to 
derive the gluon structure functions have also been developed \cite{leerajeev}.
We hope to solve that problem as well in the near future.

The collinear approximation to QCD has alos been studied by other papers 
\cite{brodsky}
However the  combination of these ideas with the picture of a baryon as a
 soliton  as well as the idea of a quantum hadrondynamics seem unique to our 
approach.

The ideas we introduce  range from infinite dimensional geometry
 to the phenemenology of particle physics. Indeed a whole new set  of tools
had to be developed in order to implement the new paradigm we propose.
 Some of the methods  are 
currently not associated with particle physics:  they are more closely 
related to classical mechanics and are perhaps  
 more familiar to mathematicians. But the   new  ideas of 
one generation   become the standard lore of the next. I hope 
these lecture notes will prepare a new generation of theoretical physicists
 to pursue these ideas. Prior knowledge of quantum field theory is not
 essential, but will provide perspective. In general, a higher degree of 
mathematical maturity than specialized knowledge of particle physics 
is assumed.

In the next few sections of this introduction 
we give a quick summary of the physics background 
necessary. The reader who finds this boring should skip to the next  chapter, 
after a glance at sections 1.4 and 1.5  which are essential to the argument.

\section{Deep Inelastic Scattering}

Let us consider in more detail  the scattering of an electron by  a proton.
The electron emits a virtual photon of momentum \m{q=k_f-k_i} where
\m{k_i} and \m{k_f} are the initial and final electron momenta. This
photon then interacts with the hadron producing some state \m{|X>}
which could contain many  particles: several mesons and some excited
baryons and anti-baryons. This is  an {\it inelastic} scattering
process. If we sum over all the states \m{|X>} so produced, we will
get the `inclusive' cross-section for the scattering of the
electron. Upto well--known electromagnetic effects, the cross-section
for this scattering 
is given by the matrix element\cite{cteqhandbook}
\beqs
W_{\mu\nu}(P,q)&=&{1\over 8\pi}\sum_\sigma \sum_{X}
<P\sigma|J_\mu(0)|X><X|J_\nu(0)|P\sigma>(2\pi)^4\delta^4(P+q-P_X)\cr
&=&{1\over 8\pi}\sum_\sigma\int dx e^{-iq\cdot x}<P\sigma|J_\mu(x)J_\nu(0)
|P\sigma>.
\eeqs
 
Here, \m{J_\mu(x)} is the electromagnetic current operator which is
the one  appropriate for electron scattering: the weak interactions of the
electron are much smaller in comparison.
For neutrino scattering it would instead  be the charged weak current 
(correponding to \m{W^{\pm}}) if the neutrino is converted into an
electron; and
 the neutral current (corresponding to \m{Z^0}) otherwise. 
Also,
\m{P} is the momentum of the hadron and we assume that we have averaged over 
  the spin \m{\sigma} of the hadron. (The target is usually not polarized, 
so that we must
 average over the 
values of the spin variables. More detailed measurements with
polarized targets have been made more recently; we will not consider
these for now.) 

The tensor \m{W_{\mu\nu}} is symmetric and transverse:
conservation of the electric current gives
\beq
	q^\mu W_{\mu\nu}(P,q)=0
\eeq
 Lorentz invariance and parity (which is a symmetry of the
electromagnetic interactions) imply that   the tensor
\m{W_{\mu\nu}(p,q)} has the form
\beqs
	W_{\mu\nu}(P,q)&=&\bigg[\eta_{\mu\nu}-{q_\mu q_\nu\over
q^2}\bigg]W_1(x_B,Q^2)+\\
& & 
{1\over m^2}\bigg(P_\mu-q_\mu{P\cdot q\over q^2}\bigg)\bigg(P_\nu-q_\nu{P\cdot
q\over q^2}\bigg)W_2(x_B,Q^2).
\eeqs

Being Lorentz scalars, the `structure functions' \m{W_{1,2}} can
depend only on the Lorentz scalars \m{q^2} and \m{P\cdot q} (\m{P^2}
is fixed to have the value \m{m^2}, where \m{m} is the mass of the proton).
 When an electron is scattered against a target, the
momentum \m{q} of the photon it emits is space-like.It is conventional to use
the positive number \m{Q^2=-q^2} to describe the energy of the
photon. The higher the value of \m{Q^2} the better the resolution of
our measurement of the structure of the hadron. As the other
independent variable we can take the dimensionless ratio (`Bjorken variable')
\beq
x_B={Q^2\over 2P\cdot q}\ .
\eeq
If the hadron we are studying is stable against strong decays 
(which is always the case for experimental reasons ) it will be the lightest
particle  with its quantum numbers. Then, the mass of the intermediate
state will be greater than or equal to  the mass of the target:
\beq
	(P+q)^2\geq m^2;\quad \implies \big[{1\over x_B}-1\big]\geq 0.
\eeq
This shows that the Bjorken variable takes values in the range
\beq
	1\geq x_B\geq 0.
\eeq

It is often more convenient to use another equivalent pair of
structure functions
\beq
	F_1(x_B,Q^2)=W_1(x,Q^2),\quad F_2(x_B,Q^2)={P\cdot q \over m^2}
W_2(x_B,Q^2).
\eeq

In its rest frame the hadron has a certain `size' \m{a}: it is of the 
 order of the
 the charge radius of the proton, \m{a^{-1}\sim 100 {\rm MeV}}.
If we take the limit \m{Q^2>>a^{-2}} keeping \m{x_B} fixed we are looking
deep inside the hadron: this is the region of Deep Inelastic
 Scattering. The basic idea is much like that of a microscope: to see
 inside an object of size \m{a} we need light of a  wavelength  that is small
 compared to \m{a}; or equivalently an energy for the photon that is
 large compared to \m{a^{-1}}.

When  the states are normalized by the usual convention
\beq
  <P|P'>=2E_P(2\pi)^3\delta^3(P-P'),
\eeq 
the functions \m{F_{1,2}(x_B,Q^2)} are dimensionless. It is found
experimentally that these functions are approximately  independent of
\m{Q^2}:  they depend essentially 
only on the dimensionless ratio \m{x_B}. The
simplest explanation of this phenomenon is that the proton is made of
pointlike massless constituents: the structure function would then 
 depend only on the dimensionless ratio
\m{x_B}.

 If we multiply  \m{P\cdot q} and \m{Q^2} by the same number, the structure
 functions are approximately invariant: this is the approximate symmetry of 
`scale invariance'. In the first versions of this model, the partons were 
concieved of 
as free particles. Of course they must  interact to bind into a hadron, 
but the idea was that at high enough \m{Q^2} the interaction would for some 
reason be small. This was the germ of the idea of `asymptotic freedom': that 
the constituents of the hadrons behaved like free particles at 
short distances, or high energies.

This `parton' model was subsequently derived \cite{thooftleshouches, grosswilczek,politzerprl}, \cite{politzerphysrep}  from a much
deeper fundamental theory of strong interactions, 
Quantum Chromodynamics (QCD). This is a non--abelian gauge theory
which has the unusual property that at high \m{Q^2} the coupling
constant (which measures the strength of the interaction) 
 vanishes like \m{1\over \log {Q^2\over \Lambda^2}}. 
Here \m{\Lambda} is a parameter with the dimensions of momentum which is
a fundamental parameter of QCD.
Hence at
high energies (compared to \m{\Lambda})  QCD tends to  a free theory, 
yielding asymptotic freedom and 
the parton model. In the limit of large \m{Q^2}, the structure functions are
 predicted to be independent of \m{Q^2}.
At finite values of
\m{Q^2}, the structure functions have a slow dependence on \m{Q^2}; as
a  polynomial in the coupling constant, or equivalently,
\m{1\over \log {Q^2\over \Lambda^2}}. Moreover, we can calculate this dependence on \m{1\over \log {Q^2\over \Lambda^2}} using perturbation theory: scale
invariance is broken in a way that is calculable. 
 Perturbation theory 
 allows us to calculate the \m{Q^2} dependence of
the structure functions: given the value of \m{F_{1,2}(x_B,Q^2)} at one
(large) value of \m{Q^2}, we can calculate it for any other large
value  of \m{Q^2}. ( We mean that \m{Q} should be 
 large compared to \m{\Lambda}, which is of the order of  \m{100\;\; MeV}. 
If \m{Q\over \Lambda} is not large, 
perturbatiion theory can no longer be used.)

It is one of the triumphs of modern particle
physics that perturbative QCD accurately describes these scale
violations.

However, the \m{x_B} dependence of the structure functions are not as
well-understood. Perturbation theory is not sufficient to calculate
it: \m{F_{2}(x_B,Q^2)} is in a certain sense  the probability
distribution (in momentum space) of a parton inside  a hadron.  Thus
it describes how partons bind together to form a hadron. Perturbation
theory around a free field theory  can never describe such  bound states. Due to the lack of any
fundamental understanding of the \m{x_B} dependence of the structure
functions, physicists  have been forced to extract them directly from
experiment. A whole generation of experimentalists and
phenemenologists have worked to produce a quite reliable
extraction  of  the \m{x_B} dependence of  structure functions from 
 data. It will be our goal to
explain this from the fundamental theory of strong interactions, QCD.

In the next section we will summarize the  definition of QCD as a quantum
 field theory.

\section{Quantum Chromodynamics}

Quantum Chromodynamics is the fundamental theory of strong
 interactions. It is a non-abelian gauge theory with gauge group
 \m{SU(N)}  with matter fields (quarks ) which are 
 spin half fermions transforming under ( \m{N_f} copies of )  
the fundamental representation of \m{SU(N)}. The natural number \m{N}
 is called the `number of colors'. It  has  the value  \m{3} in nature;
 but  it will be convenient to leave it unspecified  until the
 very end when we make comparisons  with data. The `number of flavors' \m{N_f}
 is \m{2} for most purposes although it can be as high as \m{6} in
 principle: the heavier flavors of quarks can be ignored for most
 purposes. Again, it is best to leave \m{N_f}  as an arbitrary parameter for now.

The action principle that defines QCD 
is
\beq
	S={N\over 4\alpha}\int \tr F_{\mu\nu}F^{\mu\nu}d^4x+
\sum_{a=1}^{N_f}\int \bar q^{a}[-i\gamma\cdot\nabla+m_a]q_a d^4x
\eeq

The Yang--Mills field strength is  
\beq
	F_{\mu\nu}=\pdr_\mu A_\nu-\pdr_\nu A_\mu+[A_\mu,A_\nu]
\eeq
where \m{A_\mu} are a set of four \m{N\times N} anti-hermitean matrices: 
\m{A_\mu dx^\mu} is a one-form taking values in the Lie algebra of the 
unitary group \m{U(N)}. In perturbation theory, this `gauge field ' 
 describes  `gluons': massless spin one particles analogous to the photon.
\m{q_a} is a fermionic field (describing quarks)  which is a Dirac spinor 
transforming in the fundamental representation of the `color' group \m{U(N)},
 for each \m{a}. The index takes values \m{a=1\cdots N_f}, where \m{N_f}
is  the number of copies of such quarks: the number of flavors.

The parameters of the theory are (in addition to \m{N} and \m{N_f}),
the quark masses \m{m_a} for \m{a=1,\cdots N_f} and the dimensionless
`coupling constant' \m{\alpha}. In fact the quantum theory is defined
by an additional  set of rules for `renormalization' which replace the constant
\m{\alpha} by a new parameter \m{\Lambda} which has the
dimensions of mass as well. These issues are well-known and have been
reviewed in several articles \cite{politzerphysrep}, so we will avoid them
here. T
he basic point is that the the coupling constant acquires a dependence on the energy scale \m{\alpha\sim {1\over \log{Q^2\over \Lambda^2}}}. 

 These \m{N_f+1} parameters should in
principle determine the masses and decay rates as well as the
structure functions of all the hadrons. It has turned out to be quite
difficult to predict hadronic properties from this fundamental
theory. We need to develop approximation methods which make the problem 
 tractable.
For the most part this problem is still not solved. In these talks we
will describe how we solved a  part of this puzzle: that of
calculating the hadronic structure functions from QCD.

\section{ Structure Functions from QCD}

We noted earlier that since the electromagnetic interactions are
well-understood, it is sufficient to concentrate on the `unknown' part
of the problem, which is  encoded in the structure functions
\m{W_{1,2}(x_B,Q^2)}. These were defined in terms of the expectation 
values  of the product of current operators. Now, part of QCD
is tractable by perturbation theory; this  allows us to reduce the
`unknown' part of the problem further to the expectation values of
simpler operators: the structure functions can be expressed in terms
of `parton distribution functions'. The problem we will solve is that 
of  calculating these distribution functions.

 The basic result that makes this reduction possible is called the
`factorization theorem': the
structure function can be written as a convolution of two factors:
one  whose
\m{x_B} and \m{Q^2} dependence can be calculated in perturbation
theory (the `hard' factor); and another (the `soft' factor) which has
an  unknown \m{x_B} dependence but which is independent of \m{Q^2}. 
 This  is a result of quantum field theory that can be proved to all
orders of perturbation theory; so it is  comparable in depth
 to the proof of renormalizability of gauge theories. We will just
give an intuitive argument,  pointing the reader to the literature on
 perturbative QCD  for details \cite{cteqhandbook, factorization}.

The  photon is scattered by a quark of charge
\m{e_a} inside the hadron. To leading order, 
the probability of the photon being
scattered by the proton can be written as the product of two pieces: 
the probablity
that  a quark of some momentum \m{k} will scatter the photon and the
probability that there is such a quark inside the proton. (Then we of
course sum over the intermediate momentum \m{k}.) This is the simplest
version of the factorization theorem. In the  next order of
perturbation theory,  we
will have to include the possibility that quark might emit a gluon
before absorbing  the  photon and then re-emit it. We can imagine the
scattering of the photon by the quark as a subprocess with its own
structure function, except that it can be calculated within
perturbation theory. The probability that a hadron contains a quark of
a given momentum is  of course not calculable in perturbation theory:
that is the distribution function which we will attempt to understand.

In other words, the structure function of the proton can be written as the
convolution of  two
pieces: the structure function of a parton (which is computable
perturbatively) and the distribution function of the parton inside the
hadron
 (which is non-perturbative).

In more detail,
\beq
	F_2(x_B,Q^2)=\sum_{i=a,\bar a,G}\int_0^1 {d\xi\over
\xi}C_2^i({x\over \xi},Q^2)\phi_{i}(\xi).
\eeq
Here \m{\phi_a(x_B)} is the probability of finding a quark of flavor
and momentum fraction \m{x_B} inside a proton; \m{\phi_{\bar a}} is
the probability for an anti-quark of flavor \m{a}, and \m{\phi_G}
for gluons. Note that all the \m{Q^2} dependence is in the first  factor; it can be
calculated as a power series in \m{{1\over \log{Q^2\over \Lambda^2}}}
using the standard rules of perturbation theory.
To leading order in perturbation theory, even the first factor is independent
 of \m{Q^2}: 
\beq
	C_2^{a}(x_B,Q^2)=C_2^{\bar a}(x_B,Q^2)=e_a^2\delta(x_B-1),
 		\quad C_2^G(x_B,Q^2)=0.
\eeq

 If \m{b_i(k),b_i^\dag(k)}
are the creation-annihilation operators for the parton, we have
\beq
	\phi_i(x_B)=\sum_\sigma\int {d^2{\bf k}_T\over
(2\pi)^2}<P\sigma|b_i^{\dag}(x_BP,{\bf k}_T)b_i(x_BP,{\bf k}_T)|P\sigma>.
\eeq
Here, we are averaging over all possible values of the part of the momentum
orthogonal  to the plane spanned by \m{P} and \m{q},
(called the `transverse momentum \m{{\bf k}_T}).

In terms of the field operators, we get the quark distribution functions
\beq
	\phi_{a}(x_B)=\half\sum_\sigma\int dy e^{-ix_BPy}<P\sigma|\bar q^a(y,{\bf 0}_T)\gamma_{-}q_a(0)|P\sigma>
\eeq
 and the anti-quark distribution functions 
\beq
	\phi_{\bar a}(x_B)=-\half\sum_\sigma\int dy
 e^{-ix_BPy}<P\sigma|q^a_\alpha(y,{\bf 0}_T)[\gamma_{-}]_{\beta\alpha}\bar q^a_\beta(0)|P\sigma>.
\eeq
The position arguments of the field operators are separated by a null
line in the plane spanned by \m{P} and \m{q}. The averaging over transverse momenta implies that in position
space, these field operators have the same transverse co-ordinates.

Since the gluons do  not carry electric charge, to the leading  order of 
perturbative QCD, the gluon distribution functions are not necessary in order 
to  understand the structure of the proton. We will 
  ignore them for 
now.

These formulae are not gauge-invariant and are to be understood in the
null gauge \m{A_-=0}. In a general gauge we should insert a parallel transport
operator \m{Pe^{\int A_-(y,{\bf 0}_T)}dy} along the null line
connecting two field operators. 

\section{ Reduction of QCD to Two Dimensions}

 We are
probing the hadron (whose momentum is a time-like vector \m{P})
 by a photon (or \m{W,Z}-boson)  with a space--like momentum
\m{q}. These two vectors define a two-dimensional subspace of
Minkowski space. There is also a space--like vector \m{a} which
characterizes the `size' of the hadron: the length of \m{a} is the 
charge radius of the hadron. In the Deep Inelastic limit,
 \m{Q^2=-q^2>>|a|^{-1}}
keeping \m{x_B={Q^2\over 2P\cdot q}} fixed. This means that the hadron
is of very large size in the directions transverse to the two
dimensional subspace spanned by \m{P} and \m{q}. By the uncertainty
principle, the transverse momentum   of its 
constituents is of order \m{|a|^{-1}}, which is thus small compared to
\m{Q}. A reasonable first approximation would be to let the hadron
have infinite extent in the transverse directions; and  to require  the momenta
of the constituents to lie entirely in the plane spanned by
\m{P,q}. In other words the approximation is that the fields are
independent of the transverse spatial co-ordinates.

This is analogous to  the procedure of dimensional reduction popular in unified field 
theories of  gravity. The main difference is that the dimensions that are
ignored do not form a  subpace of small volume: instead they are
infinite in extent. The point is that the momenta in these directions
are small which is the same as requiring that the fields are
independent of those spatial directions: a dimensional reduction.

Thus Deep Inelastic Scattering is described by the dimensional
reduction of QCD to two dimensions. That is what makes this phenomenon
accessible: two dimensional gauge theories have been understood by the
large \m{N} method. With some further (less drastic) approximations we
will be able to determine the spectrum and structure of hadrons in two
dimensions. These should then be verifiable experimentally.

The dimensional reduction of  QCD to two dimensions is given by the
action principle
\beqs 
	S&=& {N\over {4\alpha_1}}\int \tr F_{\mu\nu}F^{\mu\nu}d^2x+
\sum_{a=1}^{N_f}\int \bar
q^{a\alpha}[-i\gamma\cdot\nabla+m_a]q_{a\alpha} d^4x\cr\nonumber
& & +{N\over 2\alpha_1}\int \tr (\nabla_\mu\phi_A)^2d^2x+{N\over 2\alpha_1}\int
\tr[\phi_3,\phi_4]^2d^2x+\int \bar q^{a\alpha}(-i)\Gamma_
{ \alpha}^{A\beta}\phi_A q_{a\beta}d^2x.
\eeqs
Here we allow the indices \m{\mu,\nu} to take only the values \m{0,1}. The
gauge field now splits into a  1-form in the two dimensional space
with components \m{A_\mu} and a pair of scalar fields \m{\phi_3=A_3,
\phi_4=A_4}  corresponding to the transverse polarization states of
the gluon.

The four dimensional Dirac spinor \m{q} splits into a pair of
two--dimensional spinors \m{q^\alpha}, corresponding the two
eigenvalues of \m{\gamma_3\gamma_4}. Moreover \m{\alpha_1} is the
coupling constant of the two-dimensional theory which has dimensions
of (mass$)^2$. It is a combination of \m{\alpha} and the size of the
hadron: \m{\alpha_1\sim \alpha a^{-2}}. Thus the reduction of QCD to
two dimensions gives a gauge theory with scalar fields in the adjoint
representation, and twice as many flavors of quarks as the original theory.

We will study for the most part two dimensional QCD, defined by the
action
\beqs 
	S&=& {N\over {4\alpha_1}}\int \tr F_{\mu\nu}F^{\mu\nu}d^2x+
\sum_{a=1}^{N_f}\int \bar
q^{a\alpha}[-i\gamma\cdot\nabla+m_a]q_{a\alpha} d^4x.
\eeqs
 This is not quite the same as the dimensional reduction of {\it four}
dimensional QCD to two dimensions: there are no scalar fields. This
truncated theory will be sufficient to to determine the quark and
anti-quark structure functions to the accuracy we need.  The essential
techniques required to solve the theory including the scalar fields
have been developed as well. We will return to them in a later
publication.

\section{Two Dimensional QCD in Null Gauge}

The essential simplification of two dimensional QCD is that the gauge
fields can be removed completely from the problem, leaving just the
quark fields \m{q} and \m{\bar q} and the scalar gluons \m{\phi_A}.
This should not be too surprising: in \m{D} dimensional space-time, a
gauge field has \m{D-2} polarization states.( Of course, we still have 
the scalar fields \m{\phi} which are the remnants in two dimensions of the 
two polarization states of the gluon in 
four dimensional gauge theory. But we will 
ignore them, as they are mostly  relevant to the determination of the gluon 
structure functions: a problem we will postpone to a later publication.) 

We will change notation slightly from the last section:the indices
\m{a,b} etc. will denote the pairs \m{a\alpha}, \m{b\beta}, so they
will range over \m{1,2,\cdots 2N_f}.This simply reflects the fact that
the two dimensional theory has twice as many flavors as the four
dimensional theory: the transverse polarization label looks just just
like a flavor index to the two-dimensional theory.

In the null  gauge, the action of the theory becomes (see  Appendix A)
\beq
	L=\chi^{\dag a }(-i\pdr_t)\chi_{a}-
\chi^{\dag a }\bigg\{\half\big[\hat p +{m^2\over \hat p}\big]-
iA_t\bigg\}\chi_{a}+{N\over 2\alpha_1}\tr[\pdr_xA_t]^2.
\eeq

The field \m{A_t} carries no dynamical degrees of freedom: it can be
eliminated in terms of \m{\chi} by solving its equations of motion:
\beq
	A_{t j}^i(x)={i\over N}\alpha_1\int \half|x-y|
:\chi^{\dag a i}(y)\chi_{ a j}(y):dy.
\eeq
(The Coulomb potential in one-dimensional space is given by
\m{ {\pdr^2\over \pdr x^2}\half|x-y|=\delta(x-y)}.)
The hamiltonian of the resulting theory is
\beqs
	H&=&\int dx\chi^{\dag a  i}  \half[\hat p+{m^2\over \hat p}]   \chi_{a i }\cr
& & -{1\over 2N}\alpha_1\int \half |x-y|
:\chi^{\dag a i}(x)\chi_{ a j}(x):
:\chi^{\dag b j}(y)\chi_{ b i}(y):dxdy
\eeqs
Now define the operator 
\beq
	\hat M^{a}_{b}(x,y)=
-{2\over N}:\chi^{\dag a i}(x)\chi_{ b i}(y):
\eeq
which is gauge invariant. It describes the creation of a quark at \m{x}
and an anti-quark at \m{y}, but in a color invariant combination: in
other words it describes a meson. 
The hamiltonian above can be expressed
entirely in terms of this operator after some reordering of factors of
\m{\chi}.

We can  rearrange the quartic operator in the \m{\chi}'s as a quadratic
 operator in \m{M}. First,
\beqs
:\chi^{\dag a i}(x)\chi_{ a j}(x)::\chi^{\dag b j}(y)\chi_{ b i}(y):&=&
:\chi^{\dag a i}(x)\chi_{ a j}(x) \chi^{\dag b j}(y)\chi_{ b i}(y):\cr
& &+:\chi^{\dag a i}(x)\chi_{ b i}(y):<0|\chi_{ a j}(x)\chi^{\dag b j}(y)|0>\cr
& & +
<0|\chi^{\dag a i}(x)\chi_{ b i}(y)|0>:\chi_{ a j}(x)\chi^{\dag b j}(y):
\eeqs
Now,(see Appendix A)
\beq
<0|\chi^{\dag a i}(x)\chi_{ b i}(y)|0>=N\delta^a_b\int_{-\infty}^0{dp\over 2\pi}e^{-ipx+ipy}=N\delta^a_b\half[\delta(x-y)+\eps(x-y)]\nonumber
\eeq	
where,
\beq
	\eps(x-y)={\cal P}\int \sgn(p)e^{ip(x-y)}{dp\over 2\pi}.\nonumber
\eeq
Thus,
\beqs
:\chi^{\dag a i}(x)\chi_{ a j}(x)::\chi^{\dag b j}(y)\chi_{ b i}(y):&=&
:\chi^{\dag a i}(x)\chi_{ a j}(x) \chi^{\dag b j}(y)\chi_{ b i}(y):\cr
& &+:\chi^{\dag a i}(x)\chi_{ a i}(y):
:{N\over 2}[\delta(x-y)+\eps(x-y)]\cr
& & +
{N\over 2}[\delta(x-y)+\eps(x-y)]:\chi_{ a i}(x)\chi^{\dag a i}(y):
\eeqs

On the other hand,
\beqs
\hat M^{a}_{b}(x,y)\hat M^{b}_{a}(y,x)&=&
\bigg({2\over N}\bigg)^2
:\chi^{\dag a i}(x)\chi_{ b i}(y)::\chi^{\dag b j}(y)\chi_{ a j}(x):\cr
&=&\bigg({2\over N}\bigg)^2:\chi^{\dag a i}(x)\chi_{ b i}(y)\chi^{\dag b j}(y)\chi_{ a j}(x):
\eeqs
The terms corresponding to other orderings of \m{\chi} will involve \m{\eps(y,y)} which should be interpreted as zero. So,
\beqs
:\chi^{\dag a i}(x)\chi_{ a j}(x)::\chi^{\dag b j}(y)\chi_{ b i}(y):&=&
-\bigg({N\over 2}\bigg)^2\hat M^a_b(x,y)\hat M^b_a(y,x)\cr
& &+\bigg(-{N\over 2}\bigg)\hat M^a_a(x,y)
{N\over 2}[\delta(x-y)+\eps(x-y)]\cr
& & +
{N\over 2}[\delta(x-y)+\eps(x-y)]\bigg(-{N\over 2}\bigg)\hat M^a_a(y,x)
\eeqs
Thus,
\beqs
	H&=& \bigg(-{N\over 2}\bigg)\int  \half\big[
p+{\tilde \mu^2\over  p}\big] \hat{\tilde M}^a_a(p,p){dp\over 2\pi}   +\cr
& & -{1\over 2N}\alpha_1\int \half |x-y|
\bigg\{
-\bigg({N\over 2}\bigg)^2\hat M^a_b(x,y)\hat M^b_a(y,x)\cr
& &+\bigg(-{N^2\over 2}\bigg)\hat M^a_a(x,y)
\eps(x-y)
\bigg\}dxdy
\eeqs
We have dropped terms involving \m{\delta(x-y)|x-y|} since this product is
 just zero; moreover the two terms involving \m{\eps(x-y)} are equal and have 
been combined.

Now we use the identity 
\beq
	\int \half|x-y|\eps(x-y)f(x-y)dxdy=-{1\over \pi}{\cal P}\int {1\over p}\tilde f(p,p){dp\over 2\pi}.
\eeq
Thus
\beqs
	{H\over N}&=&-\half\int  \half\big[
p+{\tilde \mu^2\over  p}\big] \hat{\tilde M}^a_a(p,p){dp\over 2\pi}    +\cr
& & {1\over 8}\alpha_1\int \half |x-y| \hat M^a_b(x,y)\hat M^b_a(y,x)dxdy
\eeqs
where
\beq
	\tilde \mu^2=m^2-{\alpha_1\over \pi}.
\eeq
The commutation relations can also be expressed entirely in terms of the variables \m{\hat M}:
\beqs
	\{\tilde {\hat M}^a_b(p,q),\tilde {\hat M}^c_d(r,s)\}&=&
{1\over N}\bigg(\delta_b^c 2\pi \delta(q-r)[\delta^a_d\sgn(p-s)+\tilde {\hat M}^a_d(p,s)]\cr
& &-\delta_d^a 2\pi \delta(s-p)[\delta^c_b\sgn(r-q)+\tilde {\hat M}^c_d(r,q)]\bigg). 
\eeqs	
We notice that in the limit of large \m{N}, these commutators become small:
 it is a sort of classical limit.

\section{Constraint on the  Variable \m{\hat M} }

We have all the essential ingredients of a reformulation of two
dimensional QCD in terms of the color singlet variable \m{M},
eliminating quarks and gluons from the picture. The commutation
relations and the hamiltonian together inply the time evolution
equations for \m{\hat M}. However there is one more ingredient which we
can miss at first: the set of allowed values of \m{\hat M}. Being
bilinear in the fermionic variables, \m{\hat M} behaves much like
bosonic variables: they satisfy commutation relations (rather than
anti-commutation relations). Also, they have a classical limit; in our
case this is the large \m{N} limit when there commutators become
small. But their origin as fermion bilinears have a residual effect:
they satisfy a quadratic constraint. This constraint is  ultimately an
expression of the Pauli  exclusion principle; even in the large \m{N}
limit the fact the underlying degrees of freedom are fermions cannot
be ignored.

Regard  the classical variable \m{\rho(x,y)} 
obtained by taking the large \m{N}
 limit of  \m{{1\over N}\chi^{i\dag}(x)\chi_{i}(y)} as  a 
 matrix in the variables \m{x} and \m{y}. It  is in fact the density
 matrix of quarks. Its eigenvalues have to be between zero and one: zero when
 the state is completely empty and one when it is filled. For the
 color singlet states (hadrons), each  state is either completely
 filled with quarks or completely empty. Hence the eigenvalue is
 either 0 or one, and this density matrix is a projection operator:
\m{\rho^2=\rho} or,
\beq
	\int \rho(x,y)\rho(y,z) dz=\rho(x,z).
\eeq
This is a quadratic constraint on the classical variable.

Another  (oversimplified) way to understand this constraint is that 
\m{\rho(x,y)}
corresponds to a meson state where a quark is created at \m{x} and an
anti-quark at \m{y}. If we now create two mesons, with the position of
the anti-quark of the first coinciding with the quark of the second,
they will annihilate eachother leaving us with one meson!.

We will usually use  the  normal ordered variable 
\m{M(x,y)={1\over N}:\chi^{i\dag}(x)\chi_{j}(y):}, which has the
advanatage that \m{M=0} on the vacuum. In terms of it the constraint
becomes, in matrix notation
\beq
		[\eps,M]_++M^2=0.
\eeq
\m{M} has to  satisfy some technical conditions as
well, but we will talk about them later.

It is possible to prove these constraints by a straightforward
calculation on the operators, even when \m{N} is held finite. ]
(See the appendix of Ref. \cite{sphqcd}.)
 It is important to note that they only
hold in the color singlet subspace of the fermionic operators.

This means that it is possible to understand two dimensional QCD
without ever mentioning quarks, antiquarks or gluons. The operator
\m{\hat M} is gauge invariant and can be the field variable of a
theory that {\it directly} describes hadrons. From this point of of
view quarks and gluons are just mathematical artifacts. This gives us
a whole new paradigm for the theory of strong interactions.
Instead of  a 'Quantum ChromoDynamics'' in terms of unobservable
quarks and gluons 
 but a new ``Quantum HadronDynamics'' in terms of the directly
observable particles, the  hadrons.

In the next
chapter we will develop just such a point of view. 
We will first present a classical theory which is equivalent to the
large \m{N} limit of the above theory. Upon quantization we will
recover two-dimensional QCD.

\chapter{ Two-Dimensional Quantum Hadron Dynamics}

In this chapter we will present first the classical hadron theory and then its quantization. We will see that the quantum hadron theory is equivalent to QCD in two dimensions. 

\section{Grassmannians}

Let \m{{\cal H}=L^{2}(R)\otimes C^{N_{2f}}} be the  
complex Hilbert space of complex-valued  functions on the real
line\footnote{The number \m{N_{2f}} is the number of flavors of the two
dimensional theory; it is twice the
number of flavors of the four-dimensional theory.}. Define the
subspace of  functions whose Fourier transforms vanish for negative momenta:
\beq
	{\cal
H}_+=\bigg\{\psi|\psi(x)=\int_0^\infty\tilde\psi(p)e^{ipx}{dp\over 2\pi}\bigg\}
\eeq
and its orthogonal complement
\beq
	{\cal
H}_-=\bigg\{\psi|\psi(x)=\int_{-\infty}^0\tilde\psi(p)e^{ipx}{dp\over 2\pi}\bigg\}.
\eeq	
The functions in \m{{\cal H}_+} are boundary values of analytic
functions on the upper half of the complex plane in the variable
\m{x}. (Of course \m{{\cal
H}_-} is related  to the lower half of the complex plane in the same
way.) Define the operator \m{\eps} (`the sign of the momentum') 
to be \m{-1} on \m{{\cal H}_-} and
\m{+1} on \m{{\cal H}_+}. Clearly
\beq
	\eps^{\dag}=\eps,\quad \eps^2=1.
\eeq

The Grassmannian of \m{{\cal H}} is the set of all its subspaces. 
To each subspace \m{W},  there is an  operator \m{\Phi} which is \m{-1} on
\m{W} and \m{+1} on \m{W^\perp}. Again,
\beq
	\Phi^{\dag}=\Phi,\quad \Phi^2=1.
\eeq
Conversely any such operator corresponds to an orthogonal splitting of
\m{{\cal H}}: it will have eigenvalues \m{\pm 1} and \m{W} can be
identified as the subspace with negative eigenvalue. Thus the set of
such \m{\Phi}'s is the same as the set of subspaces of \m{\cal
H}.The set of subspaces of \m{\cal H}  is the Grassmannian, which we
can thus also regard as the set of operators \m{\Phi} satisfying 
the above condition.

But we dont want to allow all such subspaces: since \m{\cal H} is
infinite dimensional, that would give a Grassmannian that is too
big. For example, its tangent space will not be a Hilbert space, and
it wouldnt admit a Riemann metric. We should  only allow subspaces \m{W}
 that are `at a finite distance' from \m{{\cal H}_-}. 
It is convenient to introduce the variable 
\beq
	M=\Phi-\eps
\eeq
which measures the deviation from the standard point \m{\eps}, which
 corresponds to the subspace \m{{\cal H}_-}. We will
require that \m{M} be Hilbert-Schmidt. That is, the sum of the absolute
magnitudes squared of all its matrix elements is finite. (See Appendix B for a 
rapid summary of some  functional analysis we will need.)
This sum is a measure of the distance of \m{W} from \m{{\cal H}_-}.

Thus we define the restricted Grassmannian to be,
\beq
	\Gr({\cal H},\eps)=\{M^{\dag}=M,\quad [\eps, M]_+ +M^2=0, \tr M^2<\infty\}.
\eeq
The set of all such \m{M}'s forms an infinite dimensional Hilbert
manifold. This manifold
 is defined by a quadratic equation in the
(real) Hilbert space of self-adjoint Hilbert-Schmidt operators.

If we split the operator into \m{2\times 2} blocks 
(as in \m{\eps=\pmatrix{-1&0\cr
		         0&1}}),
\beq
	M=\pmatrix{a&b\cr
	           b^{\dag}&d}
\eeq
we get (using the constraint on \m{M}),
\beq
	a=\half(bb^{\dag}+a^2),\quad d=-\half(b^{\dag}b+d^2),\quad ab+bd=0.
\eeq
Since \m{M} as a whole is Hilbert-Schmidt,
 \m{b:{\cal H}_+\to {\cal H}_-} is Hilbert-Schmidt as well. The above
constraints then imply that \m{a} and \m{d} are trace-class. This
means roughly speaking that the sum of their eigenvalues is
absolutely convergent (see the appendix B  for precise definition).

Note that \m{a} is a positive operator and \m{d} a negative operator. In particular, \m{\tr a\geq 0} and \m{\tr d\leq 0}.

 The tangent space at the origin is given by the special case where
\m{a,b,c,d} are all infinitesimally small.
 The constraints then show that  \m{a} and \m{d} are second order
infinitesimals while  \m{b} are first order.
  In the first order  the only constraint on  \m{b} is that it 
has to be a  Hilbert-Schmidt
operator. Thus, the
tangent space at the origin is a complex Hilbert space, consisting of
 operators of the form \m{M=\pmatrix{0&b\cr b^{\dag}&0}}. In other words we
 have the identification,
\beq
	T_0\Gr({\cal H},\eps)={\cal I}^2({\cal H}_+\to {\cal H}_-).
\eeq

\section{The Dirac Theory of Fermions}

We have introduced the Grassmannian  from a purely geometric point of
view above. But it has  a natural physical interpretation in terms of
the Dirac theory of  fermions.

In Dirac's theory, the complex Hilbert space  \m{\cal H} represents
the set of states of a single fermion. There is a self-adjoint
operator \m{h} representing energy. In the null co-ordinates we are
using, energy is
\beq
	p_0=\half[p+{m^2\over p}].
\eeq
Thus it has the same sign as the null component of momentum
\m{p}.  The main physical obstacle to this interpretation is that the
energy is not positive: all states are thus unstable with respect to
decay into the negative energy states of lower and lower energy.

Dirac's main idea was that in any physical state all except a finite
number of the negative energy states are occupied. Since fermions obey
the Pauli exclusion principle, only one  particle can occupy a given
state: decay into states of very low energy is forbidden because
those states are occupied. In particular the vacuum is not the state
containing no particles. Instead the vacuum is the state where all the
negative energy states are occupied and the positive energy states are
empty. The filled negative energy states in the vacuum is called the
`Dirac Sea'. All physical quantities such as energy or charge are to
be measured in terms of their departure from the vacuum value: even
though the vacuum contains an infinite number of particles, it is
still assiged zero energy, charge etc.

 An excitation from the vacuum could be a state where a finite
number of positive energy states are occupied and a finite number of
negative energy states are empty. Such a state will have
positive energy compared to the vacuum. The empty negative energy
states (also called `holes') have positive energy. The holes
themselves behave just like the particles except that they have  the
opposite value for some observables such as electric charge. They are
the `anti-particles'. The number of particles minus the anti-particles
is a conserved quantity which can take any integer value.
An arbitrary physical state is
a linear combination of such states containing a finite number of
particles  and anti-particles.

A mathematical interpretation of this situation can be given in terms
of a modified  exterior product of \m{{\cal H}}. In familiar
non-relativistic theories, the space of states of a multi-particle
system of fermions would be  the exterior power of order \m{r}, 
where  is the number of fermions:
\m{\sum_{r=0}^\infty \Lambda^r{\cal
H}}.  Instead in Dirac's theory we split 
the one particle Hilbert space
into two subspaces \m{{\cal H}_-\oplus {\cal H}_+} of negative and
positive energy. The space of states of the multi-particle system  is
\beq
	{\cal F}=\sum_{r,s=0}^\infty \Lambda^{s}{\cal H}_-' \otimes
\Lambda^{r}{\cal H}_+.
\eeq
This space of multi-particle states is called the Fock space.

There is a subset of states in this Fock space cosnisting of  wedge
products of single particle states. ( A general state is a linear
combination of such wedge products.) Suppose,   \m{W} is a subspace
which doesnt differ `too much' from \m{{\cal H}_-}. That is, the
intersection  of \m{W} with \m{{\cal H}_+} 
and the intersection of \m{W^\perp} with \m{{\cal H}_-} are both finite
dimensional. Then we can form a state in \m{\cal F}  in which all
states of \m{W} are occupied and all those in \m{W^\perp} are empty.
Suppose  \m{e_i} is a basis in  \m{W\cap
{\cal H}_+}  and  and \m{f_j} a basis in \m{W^\perp\cap{\cal
H}_-}. Then the state in \m{\cal F} corresponding to occupying \m{W}
will contain particles in \m{W\cap
{\cal H}_+} and holes in \m{W^\perp\cap{\cal
H}_-}:
\beq
f^{1'}\wedge f^{2'}\wedge \cdots \wedge e_1\wedge e_2\wedge e_3\wedge\cdots.
\eeq
(Here, \m{f^{j'}} is the dual basis in \m{\big[W^\perp\cup{\cal
H}_-\big]'}).
Moreover this is independent (upto multiplication by a complex number)
of the choice of basis. Thus to each such subspace \m{W\subset {\cal
H}}  there is a state (upto scalar multiple) in the Fock space \m{\cal F}. 
In other words we have an embedding of the Grassmannian into the Projective
 space \m{{\cal P}{\cal F})}. This is the infinite dimensionl version of the
 Pl\"ucker embedding familiar from algebraic geometry. \cite{Chern,Mickelsson}.

The Grassmannian describes this subset of states  in Dirac's
theory. The condition that the operator \m{M=\Psi-\eps} corresponding
to a subspace \m{W} be Hilbert-Schmidt is precisely what is required for the
above construction to go through. (It is not necessary to require that the
projection operator \m{\pi_+:W\to {\cal H}_+} be finite dimensional as
we did above: the precise condtion for the construction to work is
that it  be Hilbert-Schmidt.) What is special about this family of
states is that they are coherent states: they  minimize the
uncertainty in the physical observables. Hence they are the states
that have a sensible classical limit. The set of such coherent states
is the classical phase space of the theory. That is why it is sensible
to choose the Grassmannian as the classical phase space of our theory
of hadrons.

Dirac's          theory was originally meant to describe electrons. We should
think of it now as applied to quarks. Since the quarks are also
fermions, this is reasonable. But there is one impotant twist: each
quark comes in \m{N} colors: thus each state in \m{\cal H} can be
occupied by \m{N} quarks, not just one. Moreover, only states that are
invariant under the action of the group \m{SU(N)} are allowed: these are the 
states that describe  the hadrons.  The classical limit referred to above is
the large \m{N} limit.

The operator \m{\Phi} has eigenvalue \m{-1} on states that are completely 
filled and \m{+1} on states that are empty. The variable \m{M}
measures the deviation from the vacuum state, in which all the
negative energy states are completely filled.
A baryon is essentially a color invariant bound state of \m{N}
quarks. Thus the Dirac Sea of quarks can also be thought as a sea of
baryons: each state is filled either by \m{N} quarks or by a
baryon. An infinitesimal
disturbance from this can be thought of either as the creation of a
meson ( a quark--anti-quark pair) or as the promotion of a quark from
the negative energy sea to a positive energy state.  This is an
infinitesimal change if the number of colors \m{N} is large. The
operator \m{M} thus has to be determined by 
a map  \m{b^{\dag}:{\cal H}_-\to {\cal H_+}}. The block-diagonal  elements of
\m{M} are zero in this infinitesimal limit.

In addition to such small deviations from the vacuum, the theory also allows 
for topological solitons which are not connected to the vacuum by any
 continuos path. These are very important, as they describe the baryons. To 
see how they arise we need to understand the topological properties
 of the Grassmannian.  This is the subject of the next section.
 
\section{Renormalized Dimension of a Subspace}

Recall that \m{\Phi=\eps+M} has eigenvalues \m{\pm 1}. Hence in some formal 
sense the trace of \m{\Phi} is an even integer:
 the difference between the number of positive eigenvalues and the number 
of negative eigenvalues. 
But of course this trace is
not  convergent. Even if we subtract the contribution of the vacuum, it
is divergent: the  \m{\tr M} can diverge, only the trace of its square
needs to be convergent in general, since we only required it to be
 Hilbert-Schmidt.
 However the trace of \m{M} is
conditionally convergent: we can define it to be the  trace with respect to the
above splitting into submatrices:
\beq
	{\rm tr}_\eps M=\tr a+\tr d.
\eeq
This trace exists since \m{a} and \m{d} are trace-class matrices. It
can be shown  to be an even integer, which is invariant under the
continuous  deformations of
 the operator \m{M}.

The interpretation of \m{\Phi} in terms of subspaces  will help us
understand the meaning of this integer. Imagine that we take  a state
from \m{{\cal H}_+} and add it to \m{{\cal H}_-} to get a subspace
\m{W}: then \m{W} has dimension one more than \m{{\cal H}_-}. This should
change the trace of \m{\Phi} by \m{-2}: one of the eigenvalues of
\m{\Phi} has changed from \m{+1} to \m{-1}.
Since \m{\tr M}
is essentially the difference between the traces of \m{\Phi} and
\m{\eps}, we  have \m{{\rm tr}_\eps M=-2}. Thus
 \beq
B=-\half\tr M \eeq 
is the `renormalized dimension' of the subspace it describes: the
difference between its dimension and the dimension of the standard
subspace \m{{\cal H}_-}.

In fact this is the only topological invariant of \m{M}: any two
operators with the same conditional trace can be connected to each other by a
continuous path. Thus the restricted Grassmannian is a union of
connected components labelled by an integer, which is called the
`virtual rank' or `renormalized dimension' of that component. Since it is 
invariant under all
continuous deformations, in particular it will be invariant under time
evolution. This is true for any reasonable definition of time
evolution, independent of the choice of hamiltonian. Thus the renormalized 
dimension   is a `topologically conserved quantity'. 

Each connected component of the Grassmannian 
 by itself is an infinite dimensional
manifold. Although the different components are the same
(diffeomorphic) as manifolds, the component of rank zero has a special
role as it contains the `vacuum' \m{M=0}. An example with 
 renormalized dimension  one is  a `factorizable' operator,
\beq
	M=-2\psi\otimes \psi^{\dag}.
\eeq
The constraints on \m{M} are satisfied if \m{\psi} is a positive
energy state of length one:
\beq
	\eps\psi=\psi,\quad ||\psi||^2=1.
\eeq

In the interpretation in terms of the Dirac theory, we have filled a
positive energy state \m{\psi} in addition to all the negative energy
states. Hence this state contains an additional fermion occupying the
state \m{\psi}.

More generally, let \m{\psi_a} be an orthonormal system of states
satisfying
\beq
	\eps\psi=\eps_a\psi,\quad <\psi_a,\psi_b>=\delta_{b}^a.
\eeq 
Then 
\beq
	M=-2\sum_a\eps_a\psi_a\otimes \psi_a^{\dag}
\eeq
is a solution of the constraints with 
 renormalized dimension  \m{\sum_a\eps_a}.
 In
this configuration, we have filled a certain number if positive energy
states (\m{\psi_a} with positive \m{\mu_a}) while creating holes in
some others, \m{\psi_a},  with negative \m{\mu_a}. The renormalized dimension
 is
just the difference between the number of occupied positive energy
states and 
the number of holes in the negative energy states.

Thus  we see that the renormalized dimension is just the  fermion number if we
interpret the Grassmannian in terms of the Dirac Sea. 
We will see that the separable  configuration is related to the valence quark 
model of  the baryon.

As noted earlier, \m{a\geq 0} and \m{d\leq 0}; thus \m{a} contributes
negatively to the baryon number and \m{d} positively. The only
configurations that have \m{b=0} are the separable ones above. To see this note that,
\m{b=0} implies that  \m{a} and \m{d} are proportional to projection
operators. In order to be trace class, these have to be finite rank
projections: corresponding to a certain number of baryons and anti-baryons.

The valence parton approximation of
the parton model corresponds to the separable ansatz in our soliton
model.  In that approximation, \m{\phi_{\bar a}(p)} is
zero: there are no antiquarks in the proton within the valence
approximation. 

\section{Some Submanifolds of the Grassmannian}

In principle a baryon can be in any of the configurations of renormalized 
 dimension one: the ground state (proton) will be the one of least energy. 
Once we have
 determined the energy function (see below), the proton structure functions are determined by minimizing this energy over all possible configurations.
Later on,
we will describe  a method to do just that numerically: the steepest
descent method. However this is a computationally intensive and slow
method. A much faster method will be to minimize the energy over some
submanifold of the Grassmannian, which is chosen so that the minimum
in this submanifold is close to the true minimum, yet is easier to
find.

 Physical intuition plays an important role in the choice of the
ansatz of configurations explored this way. Such
restricted phase spaces play an important role in our conceptual 
understanding as well: we will see that the
theory restricted to rank one configurations of the type 
\beq
	M=-2\psi\otimes\psi^\dag,\quad \eps\psi=\psi,\quad ||\psi||^2=1
\eeq 
is just the valence parton model: we will be able to derive the parton
model from QHD this way.

Note that a change of phase \m{\psi\to e^{i\theta}\psi} does not affect \m{M}.
Indeed the set of rank one configurations is a submanifold of the Grassmannian 
diffeomorphic to \m{{\cal P}({\cal H}_+)}. If we restrict the dynamics of our
 theory to this submanifold, we get an approximate theory which we will show is equivalent to the valence parton model.

 Allowing for a slightly larger set of
configurations will give us the parton model with Sea quarks and
anti-quarks. We will describe  here the submanifold of the
Grassmannian that descibes this approximation to our theory.
We generalize the above rank one ansatz to a finite rank ansatz
\beq
	M=\mu_{\alpha\beta}\psi_\alpha\psi_\beta^\dag.
\eeq
Here, the  \m{\psi_\alpha} are a finite number \m{r} of  eigenstates of
\m{\eps} which are orthonormal:
\beq
	\eps\psi_\alpha=\eps_\alpha \psi_\alpha,\quad \psi_\alpha^\dag
\psi_\beta=\delta^{\alpha}_{\beta}.
\eeq
The constraints 
\beq
	M=M^\dag,\quad (\eps+M)^2=1,\quad -\half\tr_\eps M=1,\quad \tr M^2<\infty
\eeq
become the constraints on the \m{r\times r}  matrix \m{\mu}:
\beq
	\mu=\mu^\dag,\quad [\tilde\eps+\mu]^2=1,\quad -\half\tr \mu=1.
\eeq
Here, \m{\tilde\eps=\pmatrix{\eps_1&0&\cdots\cr
			      0&\eps_2&\cdots\cr
			      \cdot&\cdot&\cdots\cr
                               0&\cdots &\eps_r}} is a diagonal  \m{r\times r}
matrix, the restriction of \m{\eps} to the finite dimensional subspace
 spanned by the \m{\psi_\alpha}.
The condition \m{\tr M^2<\infty} is of course automatic since \m{M} is
now an operator of finite rank.

Thus, given a  set of vectors \m{\psi_\alpha} satisfying the above
conditions, we get a point in the Grassmannian of renormalized
dimension one if\m{\mu} itself belongs to a {\it finite dimensional }
Grassmannian. If there are \m{r_+} vectors \m{\psi_\alpha} with positive
momentum  and \m{r_-} with negative momentum, we have 
\beq
	\tr[\tilde\eps+\mu]=r_+-r_--2.
\eeq
The simplest solution is, \m{r_+=1, r_-=0,} so that \m{\mu} is just a
number: it then  has to be  \m{-2}. This is the rank one
solution desribed earlier, which leads to the valence parton model.

If \m{r=r_++r_-} is two, there is  no solution to the above
requirements (more precisely all solutions reduce to the rank one
solution). The next simplest possibility is  of rank three,
with \m{r_+=2,r_-=1}. Then each solution to the constraints on \m{\mu} 
determines a 
one dimensional subspace of \m{C^3}; i.e., a point in \m{CP^2}. In
other words, for each such \m{\mu},
 there is a unit vector 
\m{\zeta=\pmatrix{\zeta_-\cr \zeta_0\cr \zeta_+}\in C^3} such that 
\beq
	\tilde\eps+\mu=-1+2\zeta\otimes\zeta^\dag,\quad
||\zeta||^2=1.
\eeq
Now, there is a \m{U(1)\times U(2)} ``gauge symmetry'' in the problem:
we can rotate the two positive energy vectors \m{\psi_0} and
\m{\psi_+} into each other and
change the phase of the negative energy vector \m{\psi_-} without
changing \m{M},
provided we make the corresponding changes in \m{\mu} as well. This
\m{U(1)\times U(2)} action changes the phase of \m{\zeta_-} and rotates the
other  two components among each other. This freedom can be used to choose
\beq
\zeta=\pmatrix{\zeta_-\cr
		0\cr
		\surd[1-\zeta_-^2]},\quad \zeta_- >0.
\eeq
In summary the submanifold of rank three configurations of renormalized
 dimension one is given by 
\beqs
	M&=&-2\psi_0\otimes \psi_0^\dag +
2\zeta_-\big\{\zeta_-[\psi_-\otimes\psi_-^\dag -\psi_+\otimes\psi_+^\dag]\cr
 & & +
\surd[1-\zeta_-^2][\psi_-\otimes\psi_+^\dag +\psi_+\otimes\psi_-^\dag]\big\}
\eeqs
where \m{\psi_-,\psi_0,\psi_+} are three vectors in \m{\cal H} satisfying
\beq
	\eps\psi_-=-\psi_-,\quad  \eps\psi_0=\psi_0,\quad 
\eps\psi_+=\psi_+,
\eeq
\beq
	||\psi_-||^2=||\psi_0||^2=||\psi_+||^2=1,\quad \psi_0^\dag\psi_+=0.
\eeq       
Also,
\beq
	0\leq \zeta_-\leq 1.
\eeq
The special case \m{\zeta_-=0} reduces to the rank  one ansatz. 

The physical meaning is clear if we consider the negative momentum 
components of \m{M}. The anti-quark distribution function is
\beq
	\phi_{\bar a}(p)=\tilde M_{aa}(-p,-p)=\zeta_-^2\tilde |\psi_{-a}(p)|^2.
\eeq
Thus \m{\zeta_-^2} is the probability of finding an anti-quark inside
the baryon. Also,  \m{\zeta_-^2|\psi_{-a}(-p)|^2} is the distribution
 function of the
anti-quark of flavor \m{a}. In the same way \m{|\tilde
\psi_{0a}(p)|^2+\zeta_-^2|\tilde \psi_{+a}(p)|^2} is the distribution
function for quarks. In a loose sense \m{|\tilde\psi_{0a}(p)|^2} is
the valence quark wavefunction and \m{|\tilde\psi_{+a}(p)|^2} is the
Sea quark wavefunction.  But the splitting of the quark distribution
into a valence and a Sea distribution is rather arbitrary and has no
real physical meaning.

By using the ansatz above with a larger and larger rank we can get
 better and better approximations to the Grassmannians. But the
 expressions get quite complicated beyond the rank three ansatz. We
 will find that this rank three ansatz holds the key to understanding
 the anti-quark and sea quark disributions in the quark model.
A variational ansatz based on these configurations will provide a derivation 
of the parton model with sea quarks and anti-quarks from quantum 
hadrondynamics.

\section{Integral Kernels}

We can  express the operator \m{M} in terms of its integral kernel in
 position space\footnote{ We will often suppress the indices \m{a,b}
 etc. labelling
 the basis in \m{C^{N_{2f}}}}:
\beq
	M\psi(x)=\int M(x,y)\psi(y)dy.
\eeq
Alternately we can think in terms of its action on the momentum space
 wavefunctions:
\beq
	\tilde\psi(p)=\int \psi(x)e^{-ipx}dx,\quad \psi(x)=\int
 \tilde\psi(p)e^{ipx}{dp\over 2\pi}
\eeq
and 
\beq
	\widetilde{[M\psi]}(p)=\int \tilde M(p,q)\tilde\psi(q){dq\over 2\pi}.
\eeq
The two points of view are of course related by Fourier
 transformation:
\beq
	M(x,y)=\int \tilde M(p,q)e^{ipx-iqy}{dpdq\over (2\pi)^2}.
\eeq
The operator \m{\eps} is diagonal in momentum space:
\beq
	\widetilde{\eps\psi}(p)=\sgn(p)\tilde\psi(p).
\eeq
It can also be described in position space as a distribution:
\beq
	\eps\psi(x)=\int \eps(x-y)\psi(y)dy={i\over \pi} 
		{\cal P}\int {1\over x-y}\psi(y)dy
\eeq
In fact this is a well-known object in complex function theory: the
 Hilbert transform (except for a factor of \m{i}). It relates the real
 and imaginary parts of analytic functions.

The constraints on \m{M} become then,
\beqs
	\tilde M(p,q)&=&M^*(q,p)\cr
	[\sgn(p)+\sgn(q)]\tilde M(p,q)&+& 
		\int \tilde M(p,r)\tilde M(r,q){dr\over 2\pi}=0\cr
	\int |M(p,q)|^2{dpdq\over 2\pi}&<&\infty
\eeqs
or equivalently, position space,
\beqs
	M(x,y)&=&M^*(y,x)\cr
	\int [\eps(x,y)M(y,z)&+&M(x,y)\eps(y,z)+M(x,y)M(y,z)]dy=0\cr
	\int |M(x,y)|^2dxdy&<&\infty.
\eeqs

It is useful to see what the tangent space at the origin is
like. Since \m{M} is infinitesimally small, the constraint can be
linearized:
\beq
	[\sgn(p)+\sgn(q)]\tilde M(p,q)=0.
\eeq
Thus \m{\tilde M(p,q)} is only non-zero if \m{p} and \m{q} are of
opposite signs. In fact, the case where \m{p>0, q<0} determines the
opposite one because of the hermiticity  condition on \m{M}. 
Thus the tangent space is
just the space of square integrable functions of one positive variable
and one negative variable \m{\tilde M(p,q)}.

\section{The Infinite Dimensional Unitary group}

Given any self-adjoint operator \m{\Phi}, \m{g\Phi g^{\dag}} is
also self-adjoint. If the transformation \m{g} is unitary, 
\m{gg^{\dag}=g^{\dag}g=1}, it preserves the condition
\m{\Phi^2=1}. Indeed the action of the unitary group on such operators
is transitive: any operator satisfying \m{\Phi^{\dag}=\Phi} and
\m{\Phi^2=1}  can be taken to any other by  a unitary
transformation. In terms of the variable \m{M}, the unitary
transformation is 
\beq
	M\mapsto gMg^{\dag}+g[\eps,g^{\dag}].
\eeq

In the infinite dimensional case, we have required that \m{M} be
Hilbert-Schmidt (H-S). This means that there is a corresponding condition on
the family of allowed unitary transformations: we define the restricted
unitary group,
\beq
	U({\cal H},\eps)=\{g|gg^{\dag}=1,[\eps,g]\in {\cal I}^2\}.
\eeq
It is straightforward to verify that the H-S condition on the
commutator is preserved under the  multiplication and inverse
operations, since \m{{\cal I}^2} is an ideal in the algebra of bounded 
operators. 

Indeed the restricted Grassmannian is a homogenous space of this
restricted unitary group:
\beq
	Gr({\cal H},\eps)=U({\cal H},\eps)/U({\cal H}_+)\times U({\cal
H}_-).
\eeq
The Lie algebra of the restricted Unitary group is
\beq
	\underline{U}({\cal H},\eps)=\{u|u=-u^{\dag},[\eps,u]\in {\cal
I}^2\}.
\eeq
The infinitesimal  action on the variable \m{M} is:
\beq
	M\mapsto [u,\eps+M].
\eeq

\section{Poisson Structure}

We will regard the Grassmannian as the phase space of our dynamical
system.  The matrix elements of the operator \m{M} are then a complete
set of observables.  We will seek a set of Poisson brackets among
these variables. Fortunately there is a natural choice: there is a
unique choice that is invariant under the action of the restricted
unitary group. In fact the Grassmannian can be viewed as the
co-adjoint orbit of the (central extension of the ) unitary group. We
can regard the Poisson structure as induced by the Kirillov symplectic
form on this orbit. However, the physical arguments use the Poisson
brackets rather than the symplectic form so in thsi paper we will not
say much aboutthe sympletic form. (See Refs. \cite{2dhadron, rajeevturgut}
for an elaboration of this more geometric  point of view.)

If the Poisson brackets are invariant under the action of the group,
the infinitesimal action,
\beq
	M\to [u,\eps+M]
\eeq
would be a canonical transformation for any \m{u} satisfying 
\beq
	u=-u^{\dag}, [\eps,u]\in {\cal I}^2.
\eeq
The natural choice of a function that generates this canonical
transformation  is of the form
\beq 
	f_u=k\tr\  uM
\eeq
for some constant \m{k} . There is a technical problem: the trace may not
exist. Now remember that under the splitting \m{{\cal H}={\cal H}_-\oplus {\cal
H}_+}, 
\beq
	u=\pmatrix{\alpha&\beta\cr
		    -\beta^{\dag}&\delta}\in \pmatrix{B&{\cal I}^2\cr
	      {\cal I}^2&B}, 
\quad M=\pmatrix{a&b\cr
		  b^{\dag}&d}\in\pmatrix{{\cal I}^1&{\cal I}^2\cr
	      				        {\cal I}^2&{\cal I}^1}
\eeq
so that the conditional trace 
\beq
	\tr_\eps u M =\tr \alpha a+\tr \beta b-\tr
\beta^{\dag}b^{\dag}+\tr \delta d
\eeq
exists. Moreover its value is some imaginary number, so that the
constant \m{k} should also be purely imaginary in order that 
\beq
	f_u=k\tr_\eps uM
\eeq
be a real-valued function.

Thus we postulate the Poisson brackets
\beq
	\{f_u,M\}=[u,\eps+M].
\eeq
These imply, of course that 
\beq
	\{f_u,f_v\}=k\tr_\eps v[u,\eps+M]=-f_{[u,v]}+k\tr_\eps v[u,\eps].
\eeq
The commutation relations of the restricted unitary group are
satisfied only upto a constant term: it is the central extension of
the unitary group that acts on the Grassmannian, not the group itself.
This has been studied at great length in the book by Pressley and
Segal \cite{presseg} so we wont go too far in that direction.

We can write the Poisson brackets also in terms of the integral kernel
of the  operator \m{M} in position space:
\beqs
	k\{M^a_b(x,y),M^c_d(z,u)\}&=&
\delta_b^c\delta(y-z)[\eps^a_d(x,u)+M^a_d(x,u)]\cr
& & -
\delta_d^a\delta(x-u)[\eps^c_b(z,y)+M^c_d(z,y)],
\eeqs
or momentum space:
\beqs
	k\{\tilde M^a_b(p,q),\tilde M^c_d(r,s)\}&=&
\delta_b^c 2\pi \delta(q-r)[\delta^a_d\sgn(p-s)+\tilde M^a_d(p,s)]\cr
& &-\delta_d^a 2\pi \delta(s-p)[\delta^c_b\sgn(r-q)+\tilde M^c_d(r,q)]. 
\eeqs	

The principle of invariance under the unitary group cannot determine
the constant \m{k} in the Poisson brackets. In a sense it doesnt
matter what value we choose for it, as long as the values of all the
observables are also multiplied by \m{k}. However it will be
convenient to choose the value \m{k={i\over 2}} as we will see in the
next subsection.

Soon  we will have to calculate the Poisson bracket of  functions
\m{f(M)} which is not linear in \m{M}. The derivative of such a
function can be thought of as an operator valued function \m{f'(M)} of \m{M}:
\beq
	df=\tr f'(M)dM=\int \tilde f'^b_a(M)d\tilde
M^a_b(p,q){dpdq\over (2\pi)^2}
\eeq
Then we can use the above Poisson brackets to get
\beq
	k\{f,M\}=[f'(M),\eps+M]
\eeq\label{PB}
where the l.h.s. involves the commutator, as operators, of \m{f'(M)}
and \m{\eps+M}. We leave the proof as an exercise to the reader.

\section{Momentum}

The above Poisson brackets are invariant under translations. There must be a
 canonical transformation that implements this symmetry. Formally,
 this is given by chosing \m{u} to be the derivative
 operator with respect to position; or, \m{i} times the multiplication
 by momentum.
(Strictly speaking this is  not in the Lie algebra of the
 unitary group since it is not bounded, but let us ignore this
 technicality for the moment.) Thus, momentum  is 
\beq
	P= ik\int p\tilde M(p,p){dp\over 2\pi}.
\eeq

Now, imagine calculating this quantity for the separable configuration
of renormalized dimension  one:
\beq
	M=-2\psi\otimes \psi^{\dag}, \quad ||\psi||^2=1,\quad
\eps\psi=\psi.
\eeq
With the choice \m{k={i\over 2}} we would have,
\beq
	P=\int_0^\infty p|\tilde \psi(p)|^2 {dp\over 2\pi}.
\eeq
This has a simple physical interpretation: \m{P} is just the
expectation value  of momentum is a state with wavefunction
\m{\tilde\psi(p)}.

Note that \m{P} is always a positive function: the quadratic
constraint becomes, for \m{p=q},
\beq
2\sgn(p)\tilde M(p,p)=- 
		\int \tilde |M(p,r)|^2{dr\over 2\pi}
\eeq
so that momentum can be written as 
\beq
	P={1\over 4}\int |p|\ |\tilde M(p,r)|^2{dp\over 2\pi}{dr\over
2\pi}\geq 0.
\eeq	
This makes physical sense if we regard this as the null component of
the momentum vector.

\section{ Kinetic Energy}

For a  free 
particle of mass \m{\mu}, the time-like component of momentum (energy),
\m{p_0} is related to the spatial component \m{p_1} by the condition
\beq
	p_0^2-p_1^2=\mu^2.
\eeq
If we introduce the variables
\beq	
	p=p_0+p_1, \quad p_-=p_0-p_1
\eeq
this becomes
\beq
	(2p_0-p)p=\mu^2, \quad pp_-=\mu^2.
\eeq
Thus the kinetic energy is related to the null component of momentum
through the dispersion relation:
\beq
	p_0=\half(p+{\mu^2\over p}),\quad p_-={\mu^2\over p}.
\eeq
 Notice that in this point of view
the sign of energy \m{p_0} and of momentum \m{p} are the same: 
this will prove to be convenient when studying the structure of the
ground state of relativistic fermion theories. (See the 
Appendix A  for more details.)

Thus we will postulate the kinetic energy of our dynamical system on
the Grassmannian to be 
\beq
	K=-\half\int \half[p +{\mu^2\over p}]\tilde M(p,p){dp\over
2\pi}.
\eeq
By the same argument as for momentum we can see that 
\beq
	K={1\over 8}\int [|p|+{\mu^2\over |p|}]\ |\tilde M(p,r)|^2{dp\over
2\pi}{dr\over 2\pi}\geq 0
\eeq
If we add to this an appropriate potential energy \m{U} we will get the
hamiltonian of the system. This potential energy must transform
 like \m{p_-}, in
order that \m{(p,p_-+U)} transform like the null components of momentum.

Under Lorentz tranformations,
\beq
	p\to \lambda p,\quad p_-\to \lambda^{-1}p_-.
\eeq
The Poisson bracket be invariant under Lorentz
transformations,
\beq
	\tilde M(p,q)\mapsto \tilde M_{\lambda}(p,q)=\lambda \tilde M(\lambda p,\lambda q).
\eeq
In position space, \m{x\to \lambda^{-1}x} so that 
\beq
	M(x,y)\mapsto M_\lambda(x,y)=\lambda^{-1} M(\lambda^{-1}x,\lambda^{-1}y).
\eeq
Lorentz invariance will constrain the form of the potential energy as well.

\section{Hamiltonian}

The total hamiltonian will be a sum of kinetic energy \m{K} and
a potential energy \m{U}. While the kinetic energy is best understood in
momentum space, potential energy is best written in position space.
The simplest choice of \m{U} will be a quadratic function of
\m{M(x,y)}. (Anything simpler  will lead to linear equations of motion.)
 Thus we postulate
\beqs
	U[M]&=&{1\over 4}\int M^a_b(x,y)M^b_a(y,x)v(x,y)dxdy\\
& & +
 {1\over 4}\int M^a_a(x,x)M^b_b(y,y)v_1(x,y)dxdy.
\eeqs
This will have  the requisite invariance under the internal symmetry
\m{U(F)}. 

Lorentz invariance requires that \m{U} transform like \m{p_-} or, equivalently, like \m{1\over p} or \m{x}. Thus,
\beq
	v(\lambda^{-1}x,\lambda^{-1}y)=\lambda^{-1}v(x,y).
\eeq
Moreover, they can only depend on the difference \m{x-y} due to
translation 
invariance.
Thus
\beq
	v_1(x,y)=\half \alpha_1|x-y|,\quad  v_2(x,y)=\half \alpha_2|x-y|
\eeq
for some pair of constants \m{\alpha_1,\alpha_2}.

Thus the hamiltonian of our theory is
\beqs
	E[M]&=&-{1\over 4}\int [p+{\mu^2\over p}]\tilde M(p,p){dp\over
2\pi}\cr
  & & +{\alpha_1\over 8}\int M^a_b(x,y)M^b_a(y,x)|x-y|dxdy\cr 
  & & +{\alpha_2^2\over 8}\int
M^a_a(x,x)M^b_b(y,y)|x-y|dxdy.
\eeqs
The constants \m{m,\alpha_1,\alpha_2} have to fixed later based on experimental
data. They are the coupling constants of our theory.

We have already shown that the kinetic energy is positive. The
potential energy is manifestly positive when \m{\alpha_1,\alpha_2\geq 0} 
the way we have written it in
position space. Thus the minimum value of \m{E} is zero and it is 
attained at the point \m{M=0}.( This is in fact why
we choose to parametrize our system by the variable \m{M} and not
\m{\Phi}.) This of course lies in the connected component with renormalized
 dimension  zero. In the other components,energy will have a minimum again,
but it  is not zero. In
fact we will spend much time estimating this ground state energy.

\section{Singular Integrals}

It will be convenient for later purposes to express the hamiltonian in 
momentum space variables. This will require the use of some singular integrals.

The basic singular integral is the Cauchy Principal value
\cite{hackbush} 

\beq
	{\cal P}\int_a^b {f(p)\over p-q}{dp\over 2\pi}=\lim_{\eps\to
	0^+} \bigg[\int_{a}^{q-\eps}+\int_{q+\eps}^b\bigg]{f(p)\over p-q}{dp\over 2\pi}
\eeq
This exists whenever \m{f} is H\"older continuous
\footnote{ A function \m{f} is H\"older continuous of order \m{\nu}
(or \m{f\in C^\nu}) if \m{\lim_{p\to q} {|f(p)-f(q)|\over |p-q|^\nu}}
exists for all \m{q}.}
 of order greater
than zero at the point \m{p=q}. ( Of course we are assuming 
that \m{a<q<b}). The basic idea of the Principal value integral is
to cut-off the integral by removing a small interval of width
\m{2\eps} located \m{\it
symmetrically} about the singular point and then take the limit as
\m{\eps\to 0}. 
 (If the function vanishes at the point \m{p=q} the limit might exist
even if not taken symmetrically: then we have an `improper integral'
rather than a `singular integral'.)

If we had cut-off the integral asymmetrically, the contribution of the
	region close to the singularity would have been 
\beq
	f(q)\bigg[\int_{a}^{q-\eps_1}+\int_{q+\eps_1}^b\bigg]{1\over
	p-q}{dp\over 2\pi}\sim {1\over 2\pi}f(q)\log{\eps_1\over \eps_1}.
\eeq
This can take any value as we let \m{\eps_1} and \m{\eps_1} to go to
	zero. Requiring that \m{\eps_1=\eps_2} removes this
	ambiguity. Indeed there are many other rules that could have
	been chosen: the particular one we choose must be justified by
	physical considerations: rather like the choice of boundary
	conditions in the solution of differential equations. In problems of interest to us there is a symmetry \m{p\to -p} (ultimately due to charge conjugation invariance) which selects out the Principal Value (or the Finite Part we will define soon) as the correct prescription. 

We already saw a use of the Cauchy principal value in the definition
of the operator \m{\eps}:
\beq
	\eps\psi(x)={i\over \pi}{\cal P}\int {\psi(y)\over x-y}dy.
\eeq
The symmetric choice of regularization implicit in the Principal value
is required by the condition that \m{\eps} is hermitean.

We will often have to deal with integrals that have a worse
singularity: 
\beq
	\int_a^b {f(p)\over (p-q)^2} {dp\over 2\pi}.
\eeq
The `Hadamard Finite Part' (`part finie') of such an integral is
defined  in terms of the Cauchy Principal value:
\beq
	{\cal FP}\int{f(p)\over (p-q)^2} {dp\over 2\pi}=
{\cal P}\int  {f(p)-f(q)\over (p-q)^2}{dp\over 2\pi}.
\eeq
As long as \m{f\in C^{1}}, this will
make sense: the vanishing of the numerator on the r.h.s. makes the
integral exist as a Cauchy principal value.

Although we wont need this yet, we note for completenes 
the definition of the finite part
 integral for a finite range of integration:
\beq
{\cal FP}\int_a^b {f(p)\over (p-q)^2}{dp\over 2\pi}=
{\cal P}\int_a^b  {f(p)-f(q)\over (p-q)^2}{dp\over 2\pi}
+{f(q)\over 2\pi}\left[{1\over a-q}-{1\over b-q}\right].
\eeq

We need  the notion of a finite part integral because 
\beq 
	{\cal FP}\int {1\over p^2}e^{ipx}{dp\over 2\pi}=-\half|x-y|.
\eeq
The quantity on the right hand side is a Green's function of the
Laplace operator on the real line:
\beq
	{d^2\over dx^2}\half|x-y|=\delta(x-y).
\eeq
The choice of boundary conditions in this Green's function
corresponds to the choice of the definition of the singular integral above.

As an aside we note that 
\beq
	{\cal FP}\int {g(p)\over p^2}{dp\over 2\pi}=
\lim_{\eps\to 0^+}\bigg[\int_{-\infty}^{-\eps}+\int_{\eps}^\infty\bigg]{g(p)-g(0)\over p^2}{dp\over 2\pi}\geq 0
\eeq
if 
\beq
	g(p)\geq g(0)
\eeq
for all \m{p}.
This is useful in  checking  that the hamiltonian is positive.

\section{Hamiltonian in Momentum Space}

Now,
\beqs
	\lefteqn{\half \int M^a_b(x,y)M^b_a(y,x)|x-y|dxdy=}\\
& & -{\cal FP}\int {1\over r^2}\tilde M^a_b(p,q)\tilde M^b_a(p',q'){dpdp'dqdq'dr\over (2\pi)^5}\cr
& & \int e^{ir(x-y)+ipx-iqy+ip'y-iq'x}dxdy\cr
&=& -{\cal FP}\int {1\over r^2}\tilde M^a_b(p,q)\tilde M^b_a(q+r,p+r){dpdqdr\over (2\pi)^3}
\eeqs
Similarly,
\beqs
	\lefteqn{\half \int M^a_a(x,x)M^b_b(y,y)|x-y|dxdy}\\
& & -{\cal FP}\int {1\over r^2}\tilde M^a_b(p,q)\tilde M^b_a(p',q'){dpdp'dqdq'dr\over (2\pi)^5}\cr
& & \int e^{ir(x-y)+ipx-iqx+ip'y-iq'y}dxdy\cr
&=& -{\cal FP}\int {1\over r^2}\tilde M^a_a(p,p+r)\tilde M^b_b(p',p'-r){dpdp'dr\over (2\pi)^3}
\eeqs

Thus the hamiltonian becomes
\beqs
	E[M]&=&-{1\over 4}\int [p+{\mu^2\over p}]\tilde M(p,p){dp\over
2\pi}\cr
  & & -{\alpha_1\over 4}{\cal FP}\int {1\over r^2}\tilde M^a_b(p,q)\tilde M^b_a(q+r,p+r){dpdqdr\over (2\pi)^3}\cr 
  & & -{\alpha_2\over 4}{\cal FP}\int {1\over r^2}\tilde M^a_a(p,p+r)\tilde M^b_b(p',p'-r){dpdp'dr\over (2\pi)^3}
\eeqs
In spite of the minus signs on the r.h.s., this is in fact a positive
function of \m{\tilde M}. We already saw that the kinetic energy is
positive  whenever the constraint is satisfied.

We can rewrite the second term in a more symmetric form,
\beqs
	E[M]&=&-{1\over 4}\int [p+{\mu^2\over p}]\tilde M(p,p){dp\over
2\pi}\cr
  & & -{\alpha_1\over 4}{\cal FP}\int {1\over r^2}\tilde M^a_b(p-{r\over
2},q-{r\over 2})\tilde M^b_a(q+{r\over 2},p+{r\over 2}){dpdqdr\over (2\pi)^3}\cr 
  & & -{\alpha_2\over 4}{\cal FP}\int {1\over r^2}\tilde M^a_a(p,p+r)\tilde M^b_b(p',p'-r){dpdp'dr\over (2\pi)^3}
\eeqs
Now,
\beqs
	f(r)&=&\int \tilde M^a_b(p-{r\over
2},q-{r\over 2})\tilde M^b_a(q+{r\over 2},p+{r\over 2}){dpdq\over
(2\pi)^2}\cr 
\eeqs
satisfies
\beq
	f(r)=f^*(-r)
\eeq
so that the second term in the hamiltonian can be written as 
\beq
-{\alpha_1\over 4}{\cal FP}\int {1\over r^2} {\rm Re}\; f(r){dr\over 2\pi}.
\eeq
Moreover, by a simple use of the Schwarz inequality, we see that 
\beq
	{\rm Re}f(r)\leq f(0).
\eeq
Thus this term in the hamiltonian is positive. In the same way the
last term can also be proved to be positive.

\section{ The Equations of Motion}

From our Poisson brackets and the above hamiltonian we can derive the
equations of motion. We can regard the derivative
 \m{E'(M)={\pdr E\over \pdr M}}
 of the energy with respect
to \m{M} as an operator valued function of \m{M}:
\beq
	dE=\tr E'(M)dM=\int \tilde E'^b_a(M;q,p)d\tilde M^a_b(p,q){dp dq\over (2\pi)^2}.
\eeq
Explicitly,
\beqs 
dE&=&-{1\over 4}\int [p+{\mu^2\over p}]d\tilde M^a_a(p,p){dp\over 2\pi}\cr 
  & & -{\alpha_1\over 2}{\cal FP}\int {1\over r^2}
\tilde M^b_a(q+r,p+r){dr\over (2\pi)}d\tilde M^a_b(p,q){dpdq\over (2\pi)^2}\cr
& &  -{\alpha_2\over 2}{\cal FP}\int {1\over (q-p)^2}\tilde M^b_b(p',p'+p-q){dp'\over 2\pi}d\tilde M^a_a(p,q){dpdq\over (2\pi)^2}.
\eeqs
Thus
\beqs
	\tilde E'^b_a(M;q,p)&=&-{1\over 4} [p+{\mu^2\over p}]2\pi\delta(p-q)\delta^b_a\cr 
  & & -{\alpha_1\over 2}{\cal FP} \int  {1\over r^2}
\tilde M^b_a(q+r,p+r){dr\over (2\pi)}\cr
& &  -{\alpha_2\over 2}\delta^b_a{\cal FP} {1\over (q-p)^2}\int \tilde M^c_c(p',p'+p-q){dp'\over 2\pi}.
\eeqs

Now the equations of motion implied by our Poisson brackets are, in
 operator  notation,
\beq
	k{dM\over dt}=k\{E(M),M\}=[E'(M),\eps+M].
\eeq
(Here, \m{k} is the constant we fixed earlier to be \m{i\over 2}.)
Substituting the above formula for \m{E'(M)} will give a rather complicated 
system of integral equations as our equations of motion. 

In particular, static solutions are given by 
\beq
	[E'(M),\eps+M]=0
\eeq
The obvious solution to this equation is the vacuum,
\beq
	M=0.
\eeq
This is just the point where the hamiltonian is a minimum. Actually each connected component will have a minimum for the hamiltonian; the absolute minimum is the one with renormalized dimension  zero. We will return to the study of the minimum in the components of non-zero renormalized dimension.

\section{Linear Approximation}
If \m{M} is infinitesimally close to the vacuum value \m{M=0} we can linearize the equations of motion:
\beqs
 {i\over 2}{d\tilde M^a_b(p,q)\over dt}&=&-{1\over 2}[\tilde K(p)-\tilde K(q)]\tilde  M^a_b(p,q)\cr
& &   -{\alpha_1\over 2}{\cal FP}\int {1\over s^2}
 \big[\tilde M^a_b(p+s,r+s)2\pi \delta(r-q)\sgn(q)\cr 
& & -2\pi\delta(p-r)\sgn(p)M^a_b(r+s,q+s)\big]
{drds\over (2\pi)^2}\cr
& & -{\alpha_2\over 2}\delta^a_b{\cal FP}\int \Big[ {1\over (p-r)^2}2\pi\delta(r-q)\sgn(q) \tilde M^c_c(p',p'+r-p)\cr 
& & -{1\over (r-q)^2}2\pi\delta(p-r)\sgn(p) \tilde M^c_c(p',p'+q-r)\Big]
{dp'\over 2\pi}
{dr\over (2\pi)}\cr
\eeqs
Simplifying,
\beqs
 \nonumber\lefteqn{{i\over 2}{d\tilde M^a_b(p,q)\over dt}=}\\
& & -{1\over 2}[\tilde K(p)-\tilde K(q)]\tilde  M^a_b(p,q)\cr
& &   -{\alpha_1\over 2}{\cal FP}\int {1\over s^2}
 \big[\tilde M^a_b(p+s,q+s)\sgn(q)-\sgn(p)M^a_b(p+s,q+s)\big]
{ds\over 2\pi}\cr
& & -{\alpha_2\over 2}\delta^a_b{\cal FP}\int \Big[ {1\over (p-q)^2}\sgn(q) \tilde M^c_c(p',p'+q-p)\cr 
& & -{1\over (p-q)^2}\sgn(p) \tilde M^c_c(p',p'+q-p)\Big]
{dp'\over 2\pi}\cr
\eeqs
Here,
\beq
	\tilde K(p)=\half[p+{\mu^2\over p}].
\eeq

 Recall that the constraint on \m{M} becomes, in this linear approximation,
\beq
	[\sgn(p)+\sgn(q)]\tilde M(p,q)=0;
\eeq
i.e., that \m{p} and \m{q} have opposite signs. We can assume that \m{p>0,q<0} since the opposite case is determined  by the hermiticity condition
\beq
	\tilde M(p,q)=\tilde M^*(q,p).
\eeq
Thus,
\beqs
 {i\over 2}{d\tilde M^a_b(p,q)\over dt}&=&-{1\over 2}[\tilde K(p)-\tilde K(q)]\tilde  M^a_b(p,q)\cr
& &  + \alpha_1{\cal FP}\int {1\over s^2}
 \tilde M^a_b(p+s,q+s)
{ds\over 2\pi}\cr
& & +\alpha_2\delta^a_b{\cal FP}\int  {1\over (p-q)^2} \tilde M^c_c(p',p'+q-p) 
{dp'\over 2\pi}\cr
\eeqs
Now, under translation, \m{M(x,y)\to M(x+a,y+a)} and 
\m{\tilde M(p,q)\mapsto e^{i(p-q)a}\tilde M(p,q)}. Thus the total momentum, which is a conserved quantity, is \m{P=p-q}. We can use as independent variables
\beq
	x_B={p/P}, \quad P=p-q.
\eeq
Clearly \m{0\leq x_B\leq 1} and \m{P\geq 0}.  For a stationary solution,
\beq
	\tilde M(p,q;t)=e^{i\omega t}\chi(x_B).
\eeq
We get thus
\beqs
	\nonumber\lefteqn{\half\omega\chi(x_B)=}\\
& & {1\over 4}[2P+{\mu^2\over P x_B}+{\mu^2\over P(1-x_B)}]\chi(x_B)-
{\alpha_1\over 2\pi P}{\cal FP}\int_0^1 
{\chi(y)\over (x-y)^2}dy\\
& & +{\alpha_2\over 2\pi P}\int_0^1\chi(y)dy
\eeqs

Recalling that the mass \m{\mu'} of the meson is given in light cone coomponents by, 
\beq
	2\omega P-P^2=\mu'^2
\eeq
we get 
\beqs
	\nonumber\lefteqn{\mu'^2\chi(x_B)=}\\\nonumber
& & [{\mu^2\over  x_B}+{\mu^2\over (1-x_B)}]\chi(x_B)-{2\alpha_1\over \pi }{\cal FP}\int_0^1 {\chi(y)\over (x_B-y)^2}dy-{\alpha_2\over \pi }\int_0^1\chi(y)dy
\eeqs

If we set \m{\alpha_2=0} 
this is exactly the equation that 't Hooft obtained for the meson
spectrum of two dimensional QCD.

't Hooft obtained this result by summing over an infinite class of Feynman
 diagrams. He showed that in the usual perturbaton series of two dimensional 
QCD, the diagrams of planar topology dominate in the large \m{N} limit. He 
then found a set of integral equations that describe the sum over these 
planar diagrams. 

 't Hooft's  approach is limited by its origins in 
perturbations theory. It does  not in fact reproduce the full large \m{N} 
limit of two dimensional QCD; only the {\it linear approximation} to the large
\m{N} limit is obtained this way. The complete large \m{N} limit that we have
 constructed, by very different methods, is a nonlinear  theory, reproduces
the earlier theory of 't Hooft as just the linear approximation.

Our theory of course contains much more information: it describes all the 
interactions among the mesons, since the hamiltonian is a non-linear function 
of the co-ordinates on the Grassmannian. Indeed it is not even a polynomial
 which means that there are interaction vertices at every order in 
perturbation theory. Some of these have been determined, but our analysis 
gives all the infinite number of vertices in terms of a single constant: the 
geometry of the Grassmannian fixes all these uniquely in terms of \m{\alpha_1}.
The first such vertex was obtained by a summation of planar diagrams as well;
but this approach will quickly get bogged dpwn by combinatorial  complications.
Before our work there was no indication that it was even possible to get a 
closed formula for all the interactions among the mesons in the large \m{N} 
limit of two dimensional QCD.

The solution of this linear integral equation will give the structure functions
of mesons. Experimental data exists on the structure functiosn of mesons as
 well. It is not from Deep Inelastic Scattering, but from 
Drell-Yan scattering.  Preliminary results on the comparison 
of these structure functions with data are encouraging. However the focus of 
these lectures is the much more conceptually deep problem of obtaining the 
structure function of the proton.

In fact our quantum hadrondynamics is not only a theory of mesons but also of
 baryons.  We will show in the next chapter that there are topological solitons
in our theory that describe baryons.

\section{Quantization of HadronDynamics}

We have so far described a classical theory, whose phase space is the
 Grassmannian. The Poisson bracket relations for the \m{M(x,y)} play a
 role analogous to the canonical commutation relations. There are two
 distinct approaches to quatization of this theory:\hfill\break
\noindent
(i) Find a representation of the Lie algebra in terms of operators ona
 complex Hilbert space ( `canonical quantization', 
pursued in Ref. \cite{2dhadron}; and,
\noindent
(ii) Realize the wavefunctions as holomorphic  sections of a  line
bundle on the Grassmannians (`geometrical quantization' pursued in
Ref. \cite{rajeevturgut}.

In the first approach, we can use the complete classification 
of highest  weight unitary representation of the (central extension of
the )  infinite-dimensional unitary Lie algebra, due to Kac and
Peterson  \cite{Kac}. These 
turn out to be described by  Young tableaux just as in the finite
dimensional case, except they may have infinite depth. The question
arises which irreducible representation to choose. The quadratic
constraint selects out one class of representations, whose Young
Tableaux are rectangular, of  width \m{N} and infinite depth.
Such a representation can be realized as 
\beq
	\hat M(x,y)={1\over N}:q^{i\dag}(x)q_i(y):
\eeq
where \m{q,q^{\dag}} satisfy the canonical anti-commutation
relations. The representation is carried by the  vector space of color
invariant states in the fermionic Fock space. Thus we recover exactly
the light-cone gauged fixed version of two dimensional 
 QCD with \m{N} colors  as the quantum theory of hadrondynamics. We
have  shown that two
dimensional QCD and two dimensional hadrondynamics are {\it exactly} 
equivalent at all energies  and for all values of the number of
colors.  We have already given the  details of this argument  in
Ref.\cite{2dhadron}.

In the geometric approach, we construct a line bundle of Chern number 
 \m{N} on the Grassmannian. There is a connection on this line budle
 whose curvature is the natural symplectic form of the
 Grassmannian. The infinite dimensional Grassmannian is a complex
 manifold, just like the finite dimensional ones. Holomorphic sections
 of the line bundle exist when the Chern number is  a positive
 integer. An inner product can be established on the vector space of
 these holomorphic sections by using ideas of Segal \cite{segalcmp} or
  the measure of Pickrell \cite{pickrell}. 
 The  observables \m{M(x,y)} can be represented as operators
 on  this Hilbert space. Again the formulae we get are just the same
 as that of light-cone gauge fixed QCD. See \cite{rajeevturgut} for
 details.

In addition it is sometimes convenient to quantize a restricted
version of the theory where the phase space is some symplectic
submanifold of the Grassmannian. This will give us some insight into
the connection between our theory and the parton model. We will return
to this in the next chapter.

To summarize,we have a theory that doesnt just describe mesons to all orders in
 perturbation theory; we even have a theory of baryons. The
 topologically non-trivial solutions of the theory- the solitons-
 describe baryons. In the next chapter we will study these solitons in
 more detail and obtain a theory of the structure functions of the
 baryon.

\chapter{ Baryons as Solitons and Their Structure Functions }

We saw that our phase space is a disconnected manifold, each connected 
component being labelled by an integer
\beq
	B=-\half \tr_\eps M.
\eeq
Thus each connected component will have a minimum for the energy. These will 
describe stable static solutions of the equations of motion. They are called
 topological solitons. 

The minimum of the energy 
in each connected component with a given value of \m{B} 
desribes a stable particle, as the 
only particles with lower energy would be in a sector with a different value 
of \m{B}. It cannot decay into those states since \m{B} is a conserved 
quantity.

We saw that hadron dynamics describes, in the linearized 
 approximation, the spectrum of mesons of  two dimensional QCD. What
 does a topological soliton describe? It is an old idea of Skyrme that
 the topological solitons of a theory of hadrons are baryons. This
 idea was revived in the mid eighties \cite{syracuseskyrme,rajeevthesis} and 
shown to be consistent with
 QCD in four dimensions.

With this in mind, we should expect that the topological solitons of
our two-dimensional theory describe baryons in the Deep Inelastic
region. In the center of mass frame, a baryon will (due to Lorentz
contraction) have a thin flat shape: one that can be described within a
two dimensional theory. This offers us the possibility of solving one
of the long-standing problems of particle physics: to explain the
structure functions of hadrons as measured in Deep Inelastic
Scattering. By solving the static equations of motion, we will be able
to determine the `shape' of the soliton, which will then give us the
dependence of the hadronic structure function on the Bjorken \m{x_B}
variable. The equations we need to solve are certain singular
nonlinear integral equations. They are exactly solvable, so we will
have to resort to numerical techniques.  Indeed some ingenuity is
needed even in the numerical part of this project: `off the shelf'
methods do not work due to the singularities in the integrals.
 These
singularities do not pose any new conceptual problems however: we are
dealing with a finite quantum field theory. The basic definitions of
these singular integrals ( as dealt with an earlir section) go back to
the nineteenth century work of Cauchy and Hadamard.

The static equations of motion are
\beq
	[E'(M),\eps+M]=0
\eeq
where \m{E'(M)} is the derivative of the energy,  computed  above. It
will be most convenient to view them as integral equations for \m{\tilde
M(p.q)}, the integral kernel in momentum space.
Of course we must solve these equations  subject to the constraints
\beq
	[\eps,M]_+ +M^2=0;\quad \tr_\eps M=-2.
\eeq
The last condition picks out the sector with baryon number one.

\section{Quark Distribution  Functions from QHD}

If we split the operator \m{M} into  submatrices according to the splitting \m{{\cal H}={\cal H}_-\oplus{\cal H}_+}, 
\beq
	M=\pmatrix{a&b\cr
		   b^{\dag}&d},
\eeq
the operator \m{a:{\cal H}_-\to {\cal H}_-} is positive while \m{d:{\cal H}_+\to {\cal H}_+} is negative. 
The baryon number \m{B=-\half\tr_\eps M} 
is the sum of two terms: one from positive momenta and one from negative momenta:
\beq
	B=
-\half\sum_{a}\int_0^\infty\big[\tilde a_{aa}(-p,-p)+
\tilde d_{aa}(p,p)\big]{dp\over 2\pi}.
\eeq
The first term is always negative and the second always positive.
Thus, \m{\phi_{\bar a}(p)=\half \tilde a_{aa}(-p,-p)} 
can be thought of as the
 distribution function of anti-quarks in a hadron while 
\m{\phi_{\bar a}(p)=-\half d_{aa}(p,p)}
is the distribution function for quarks. The  quantity 
\beq
	\phi_{a}^V(p)=\phi_a(p)-\bar \phi_{\bar a}(p)
\eeq
 corresponds to what is usually called 
the valence parton distribution. Its integral over \m{p} and sum over \m{a} is equal to the baryon number.

 So the quark distribution function can be thought of as the sum of the `valence' and  `Sea' contributions:
\beq
	\phi_a(p)=\phi^V_a(p)+\phi_a^S(p).
\eeq
The `Sea' quark distribution function is thus just
the same as the anti-quark distribution  function:
\beq
	\phi^S_a(p)=\phi_{\bar a}(p).
\eeq 

 We find this terminology a bit confusing and mention it
only for purposes of comparison.  The point is that there is no physical
 meaning to the  splitting of the quark distribution function into a
valence and a sea quark contribution: only the sum has a measurable,
physical significance. The anti-quark distribution does in fact have a
physical significance. The above definition of a Sea quark
distribution is thsu completely arbitrary and is {\it not} imposed on
us by any symmetry such as charge conjugation invariance. In fact we
will find later that there is another splitting which is more  natural,
where the Sea quarks have a wavefunctions are orthogonal to the
valence quark wavefunction. This at least respects the Pauli principle
for quarks. Even that spliting is merely a matter of convenience of
interpretation: all comparisons with experiment should be in terms of
the measurable quantities \m{\phi_a(p)} and \m{\phi{\bar a}(p)}.

Actually the distribution functions are usually thought of as functions of a 
dimensionless variable \m{x_B}. If \m{P} is the total momentun (in the null direction) of the hadron, the momentum of the quark can be measured as 
a fraction of \m{P}. So we should actually write
\beq
	\phi_a(x_B)=-\half \sum_{a} \tilde d_{aa}(x_B P), \quad \phi_{\bar a}(x_B)=\half\sum_a\tilde a_{aa}(-x_BP).
\eeq
These identifications of the parton distribution functions in terms of
the diagional matrix elements of \m{\tilde M_{ab}(p,p')} can also be
seen in terms of the formula we derived in terms of QCD.

The structure functions are supposed to vanish for \m{x_B>1}. Why
should the r.h.s. of the above equation vanish for such \m{x_B}?
For large \m{N}, the structure functions will be of order 
\m{e^{-cN^{3\over 2}}} at \m{x_B=1}. So even though the wavefunction 
 doesnt vanish
at \m{x_B=1}, it is exponentially small. A more proper analysis of the
large \m{N} limit, 
taking into account semi-classical corrections will
in fact reproduce wavefunctions that vanish at \m{x_B=1}.
 The difference between these  two versions of the large \m{N} limit 
 is  like the difference between the
microcanonical ensemble and the canonical ensemble in statistical
mechanics.

\section{Rank One  Ansatz}

Before we set out to solve the above integral equations, it is useful
to consider a simpler, variational approximation to them. A solution
to the above constraints on \m{M} is the rank one (or separable) ansatz,
\beq
	M=-2\psi\otimes \psi^{\dag}.
\eeq
We showed earlier that this is a solution to the constraints with
baryon number one if \m{\psi} is a positive energy wavefunction of
length one:
\beq
	\eps\psi=\psi,\quad \sum_a\int |\tilde \psi_a(p)|^2{dp\over 2\pi}=1.
\eeq
We are not claiming that the exact solution of the static equations of
motion are of this form: only that this is a reasonable variational
ansatz for it. The physical meaning of this ansatz will be made clear
in the next section: it is equivalent to the valence quark model of
the baryon in the large \m{N} limit.

The technical advantage of this ansatz is that it solves the
constraint; more precisely it reduces it to the much simpler
constraint that \m{\psi} is of length one. Geometrically, the set of
states we are considering forms a projective space: changing \m{\psi}
by a complex number of modulus one does not change \m{M}. The
separable ansatz is an embedding of the complex projective space
\m{{CP}({\cal H}_+)} of the positive energy states into the component
of the  Grassmannian with renormalized dimension  one. Instead of minimizing the
energy on the whole Grassmannian, in this approximation, we minimize
it on this submanifold.

The parton structure functions have  a simple meaning within this approximation: the structure function
 \m{\phi_a(x_B)} is just the square of the wavefunction \m{\tilde\psi(p)} in momentum space:
\beq
	 \phi_a(x_B)=|\tilde\psi_a(x_BP)|^2.
\eeq
The anti-quark structure function is just zero in this separable approximation.

The energy on this submanifold can be easily calculated:
\beqs
	E_1(\psi)&=&\sum_a\int_0^\infty\half[p+{\mu^2\over
p}]|\tilde\psi_a(p)|^2{dp\over 2\pi}\\
& & +
{\alpha_1+\alpha_2\over 2}\sum_{ab}\int |\psi_a(x)|^2|\psi_b(y)|^2|x-y|dxdy
\eeq
where as usual,
\beq
	\psi_a(x)=\int_0^\infty \tilde\psi_a(p)e^{ipx}{dp\over 2\pi}.
\eeq
\footnote{We will find it convenient to denote
\beq
	\alpha_1+\alpha_2=\tilde g^2
\eeq
to agree with previous papers.}
Such a `positive momentum' function 
has an analytic continuation into the upper half
plane in the \m{x} variable. A simple choice would be 
\beq
	\psi_a(x)={C_a\over (x+ib)^2}
\eeq
with the location of the double pole serving as a variational
parameter. \m{C_a} is constrained 
 by the normalization condition that the length
of \m{\psi} is one. (A simple pole would have infinite kinetic energy,
so we consider the next simplest possibility.) It is easy to see that
there is a term in energy that scales like \m{b^{-1}} (coming  from
\m{p|\tilde\psi(p)|^2}) and the remaining terms scale like \m{b}. Thus
there is a minimum. This encourages us to proceed to a more accurate
determination of the wavefunction \m{\psi} that minimizes the energy.

The integral equations we have to  solve are
\beqs
	\nonumber\lefteqn{\half[p+{\mu^2\over p}]\tilde\psi_a(p)+{\cal
FP}\tilde g^2\int \tilde
V(p-q)\tilde\psi_a(q){dq\over 2\pi}= \lambda\tilde\psi_a(p)}\cr 
\tilde{V}(p)&=&-{1\over p^2}\int_0^\infty
\tilde\psi^{*a}(p+q)\tilde\psi_a(q){dq\over 2\pi}.
\eeq
subject of course to \m{\tilde\psi(p)=0} for \m{p\leq 0} and
\m{\int_0^\infty |\tilde \psi(p)|^2{dp\over 2\pi}=1}.
The second of the above integral equations is just the momentum space
version of Poisson's equation of electrostatics:
\beq
	V''(x)=-\psi^{a*}(x)\psi_a(x).
\eeq
From Gauss' law, 
\beq
	V(x)\sim \half |x|
\eeq as \m{|x|\to \infty}: it is just the electrostatic potential of a
unit charge located near the origin. Thus 
\beq
	\tilde V(q)\sim -{1\over q^2}
\eeq
as \m{|q|\to 0}. This is why the integrals are singular: the finite
part prescription is just a way of imposing the boundary condition
that \m{V(x)\sim \half |x|} at infinity in position space.
We also  note that  for a solution centered at the origin, 
\beq
	\psi(x)=\psi^*(-x)
\eeq
and \m{V(x)=V(-x)}. The boundary condition at the origin we impose is 
\beq
	V(0)=0
\eeq
which translates to 
\beq
	{\cal FP}\int \tilde V(q){dq\over 2\pi}=0.
\eeq

\section{Approximate Analytic Solution for \m{\mu=0}}

It will be useful to have an approximate analytical solution; even if
it only works for a physically uninteresting region it will help us to
validate our numerical method.

Consider the singular integral (setting \m{\tilde W(q)=q^2\tilde V(q)}):
\beqs
	{\cal FP}\int \tilde V(q)\tilde\psi_a(p+q){dq\over 2\pi}&=&{\cal
P}\int {\tilde W(q)\psi_a(p+q)-\tilde W(0)\psi_a(p)\over q^2}{dq\over 2\pi}\cr 
&=& \half{\cal P}\int 
{\tilde W(q)[\tilde\psi_a(p+q)+\tilde\psi_a(p-q)]-
2\tilde W(0)\tilde\psi_a(p)\over q^2}{dq\over 2\pi}\cr
&=&\half{\cal P}\int 
\tilde
W(q){[\tilde\psi_a(p+q)+\tilde\psi_a(p-q)-2\tilde\psi_a(p)]\over
q^2}{dq\over 2\pi}\cr 
& & +
\half\bigg[{\cal P}\int {2[\tilde W(p)-\tilde W(0)]\over q^2}{dq\over
2\pi}\bigg]
\tilde\psi_a(p)\cr
\eeqs
The first integral is 
 not singular any more. The integral in the
square brackets in the last term is in fact zero by our boundary condition:
\beq
	{\cal P}\int {2[\tilde W(p)-\tilde W(0)]\over q^2}{dq\over
2\pi}={\cal FP}\int \tilde V(q){dq\over 2\pi}=V(0)=0.
\eeq
	
Thus
\beq
	{\cal FP}\int \tilde V(q)\tilde\psi_a(p+q){dq\over 2\pi}=
\half\int 
\tilde
W(q){[\tilde\psi_a(p+q)+\tilde\psi_a(p-q)-2\tilde\psi_a(p)]\over
q^2}{dq\over 2\pi}.
\eeq
So far we havent made any approximations on this integral: just
rewritten the singular integral in a better way. The main contribution
to this integral ought to come from the neighbourhood of the point
\m{q=0}. So we should be able to approximate the quatity in the square
brackes by its leading Taylor series approximation
\beq
	{\cal FP}\int \tilde V(q)\tilde\psi_a(p+q){dq\over 2\pi}\sim
\bigg[\int q^2\tilde V(q){dq\over 2\pi}\bigg]\tilde\psi''_a(p).
\eeq	 
 Then our nonlinear singular integral
equation reduces to an ordinary differential equation:
\beq
	-b\tilde\psi_a''(p)+\half [p+{\mu^2\over
p}]\tilde\psi_a(p)&=&\lambda\tilde\psi_a(p)
\eeq
where \m{b} is determined by the self-consistency constraint
\beq
	b=-\int q^2 \tilde{V}(q){dq\over 2\pi}&=&\sum_a\bigg|\int_0^\infty
\tilde\psi^{a}(q){dq\over 2\pi}\bigg|^2.
\eeq 

This equation can be thought of as  the nonrelativistic Schr\"dinger equation
for a particle of mass \m{b^{-1}} in a linear plus Coloumb
potential. The only additional complications are that \m{b} is
determined by the above self-consistency relation and also, we have
the boundary condition \m{\tilde\psi(p)=0} for \m{p\leq 0}.The only non-linearity in the problem is in the equation determining \m{b}.

Such non-relativistic models with linear plus Coloumb potentials 
  have been used to describe mesons made of   heavy quarks
  \cite{potmodel}. But the
  physical origin of this Schr\"odinger  equation  
is completely different in our case: we get this equation in
  momentum space and not position space. The eigenvalue \m{\lambda} does not
 have the physical meaning of
  energy for us: energy is to be determined by substituting the
  solution into the formula for the hamiltonian.
 Moreover, we are studying the
  fully relativistic bound state problem. And of course we are
  studying baryons not mesons. Still it is encouraging that it is
  possible to reduce a fully relativistic bound state problem to a
  mathematical problem that is no more complicated than the
  non-relativistic case.

The special case \m{\mu=0} is particularly simple. In this extreme relativistic
 case,  the solution is an Airy function:
\beq
	\tilde \psi(p)=C{\rm Ai}\bigg({p-2\lambda\over
[2b(\alpha_1+\alpha_2)]^{1\over 3}}\bigg)
\eeq
\m{C} is fixed by normalization: \m{\int|\tilde\psi(p)|^2{dp\over 2\pi}=1}.
The eigenvalue is fixed by the continuity of the wavefunction: since
it must vanish for \m{p\leq 0} we require it to vanish at \m{p=0} as
well by continuity. Then
\beq
	\lambda=-\half \xi_1[2b(\alpha_1+\alpha_2)]^{1\over 3}
\eeq
where \m{\xi=-2.33811} is the zero of the Airy function closest to the origin.
Now we fix the constant \m{b} by putting this back into the nonlinear
 self-consistency condition. We get
\beqs
	2b&=&\tilde{g}{1\over (2\pi)^{3\over 2}}{|\int_{\xi_1}^\infty\Ai(\xi)d\xi|^3\over
\big[\int_{\xi_1}^\infty \Ai^2(\xi)d\xi\big]^{3\over 2}}\cr
\lambda&=&{|\xi_1|\over 2\surd(2\pi)}{\int_{\xi_1}^\infty
\Ai(\xi)d\xi\over \big[\int_{\xi_1}^\infty
\Ai^2(\xi)d\xi\big]^{1\over 2}}\tilde{g}\sim 0.847589 \tilde{g}.
\eeqs
Moreover
\beq
	\tilde\psi(p)=C\Ai\big(\xi_1({p\over \tilde{g}\lambda}-1)\big).
\eeq
We plot the solution so obtained below.

\begin{figure}
	\centerline{\includegraphics{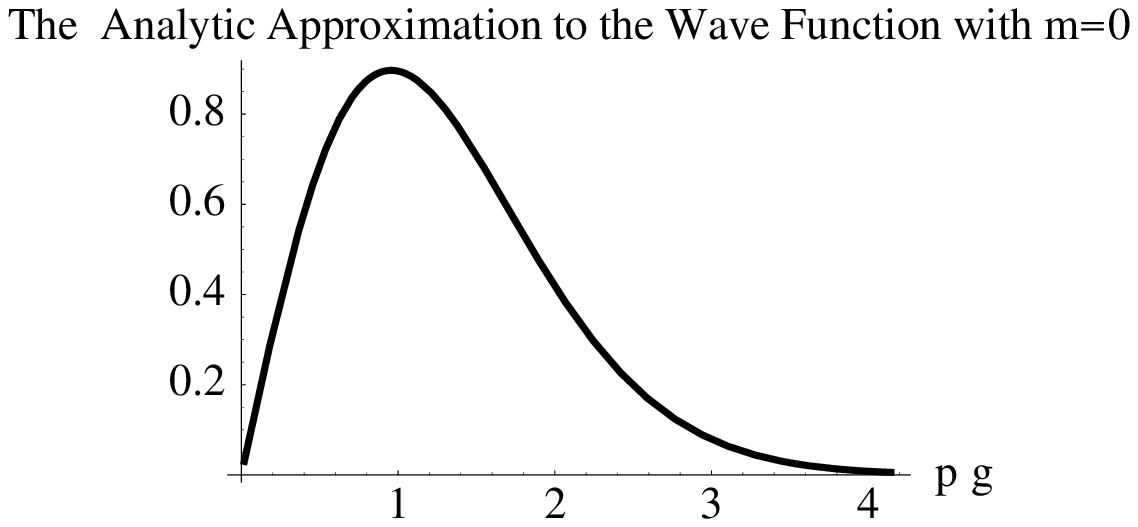}}
\caption{The approximate analytic solution for \m{\mu=0}. }
\end{figure}

\section{Numerical Solution with the  Rank One  Ansatz}

We now describe how to solve our integral equation numerically. First
we replace the nonlinear integral into a recursion relation. 
\beq
	\half[p+{\mu^2\over p}-2\lambda_s]\tilde\psi_{s+1}(p)+{\tilde
\alpha_1}{\cal P}\int_0^\infty\tilde{V_s}(p-q)\tilde\psi_{s+1}(q){dq\over 2\pi}=0
\eeq
and 
\beq
	\tilde V_s(p)=-{1\over
p^2}\int_0^\infty\tilde\psi_s^*(p+q)\tilde\psi(q){dq\over 2\pi}.
\eeq
We start with an initial guess \m{\tilde V_0(p)}, solve the linear integral
equation to get a solution \m{\tilde \psi_1(p)}. Among all the
solutions of this linear integral equation we pick the one without a
node; this happens to be the one with the smallest eigenvalue
\m{\lambda}. Then we calculate \m{\tilde V_1(p)} as above and then
again solve the integral equation to get \m{\tilde\psi_2} and so on
till our iteration  converges. 

We pick the nodeless eigenfunction at
each stage since we expect our final answer for the ground state
eigenfunction to have this property. It is just a coincidence that
this happens to have  the smallest eiganvalue \m{\lambda}: in any case
\m{\lambda} does not have the meaning of energy.

Of course to carry out this algorithm, we need to convert the above
integral equations into matrix equations. That is done by the
quadrature method (described in the appendix C ) for singular integrals. The whole
procedure can be implemented in Mathamatica quite well. 

We plot the numerical solution so obtained against the approximate
analytical solution we got earlier for the special case \m{\mu=0}.

\begin{figure}
\includegraphics{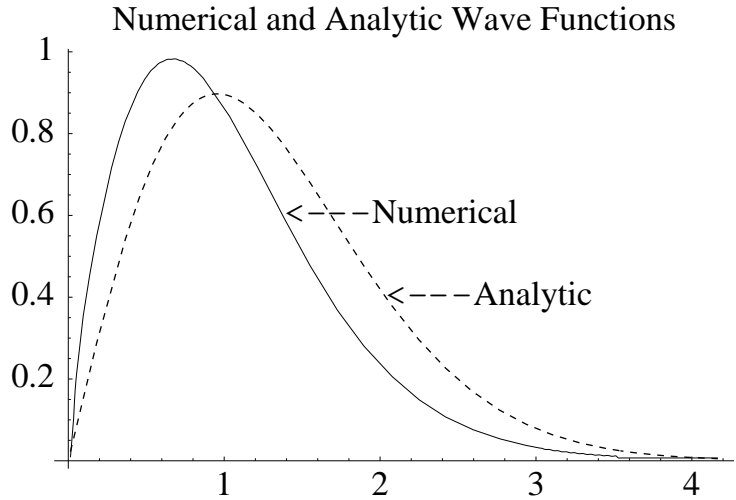}
\caption{ Comparison of Numerical and Approximate Analytic Solutions for 
\m{\mu=0}.}
\end{figure}

The curves are qualitatively the same: the difference has to do with the 
approximations we had to make in order to get an analytic solution. 
The physicaly interesting case in fact does not have \m{\mu=0}: indeed we will 
see that the value of \m{\mu^2} is in fact negative. For example when
\m{m=0}, we have, \m{\mu^2=-{\tilde g^2\over \pi}}. It is of much interest to
 see if there is a variant of the analytic approximatin method above that
 applies to this more realistic case.

\section{ Quantization of the Rank One Ansatz }

We have considered elesewhere \cite{2dhadron,rajeevturgut}  
the quantization of our hadron dynamics,
 to recover QCD.  It is also instructive to study the quantization of
 a simpler version of our  theory, corresponding to the rank one  ansatz. 
The quantization of this truncated theory gives a sort of approximation to QHD
 (hence QCD) which is of some interest in itself. We will see in the next 
chapter that this is just the  valence parton model with interactions between 
the partons as predicted by QCD.

Given a vector satisfying 
\beq
	\eps\psi=\psi,\quad ||\psi||^2=1
\eeq
we have an element of the Grassmannian of renormalized dimension  one:
\beq
	M=-2\psi\otimes\psi^{\dag}.
\eeq
But this element remains the same if we change \m{\psi} by a complex
number of modulus one: \m{\psi\to e^{i\theta}\psi}. Thus we have an
embedding of the projective space \m{{\cal P}({\cal H}_+)} into the
Grassmannian. This is a sympletic embedding: the symplectic form on
the projective space induced by this embedding is the same as the
natural sympletic form on the projective space. Thus we can regard the
separable operators as forming a `reduced phase space' describing part
of the degrees of freedom of our theory.

A way to understand this is to consider \m{\tilde\psi(p)} as a
complex-valued observable on the effective phase space. In order to
reproduce the Poisson brackets of \m{M}, these must satisfy
\beq
	\{\tilde\psi_a(p),\tilde\psi_b(p')\}=0=
\{\tilde\psi^{*a}(p),\tilde\psi^{*b}(p')\},
\eeq
\beq
\{\tilde\psi_a(p),\tilde\psi^{*b}(p')\}=-i2\pi\delta(p-q)\delta^b_a.
\eeq
They satisfy the constraints,
\beq
	\tilde\psi(p)=0;{\rm for}\; p<0,\quad \int_0^\infty |\psi(p)|^2 {dp\over 2\pi}=1.
\eeq
We should regard the function \m{\tilde\psi(p)} on momentum space to be the
 fundametal variable from which other observables such as
\beq
	\psi(x)=\int_0^\infty \tilde\psi(p)e^{ipx}{dp\over 2\pi}
\eeq
can be obtained.

Both the Poisson brackets and the constraint are simpler than the full
theory, which is why we consider this case first.

The Hamiltonian of our reduced dynamical system is obtained by putting
the ansatz into \m{E(M)}:
\beqs
	E_1(\psi)&=&\sum_a\int_0^\infty\half[p+{\mu^2\over
p}]|\tilde\psi_a(p)|^2{dp\over 2\pi}\\
& & +
{\alpha_1+\alpha_2\over 2}\sum_{ab}\int |\psi_a(x)|^2|\psi_b(y)|^2|x-y|dxdy.
\eeqs
These Poisson brackets and hamiltonian form a perfectly well--defined
dynamical system on its own right. We will now consider how to
quantize this theory, and obtain the rules for the semi-classical approximation.

We can quantize the theory by looking for operators satisfying canonical commutation relations:
\beq
	[\hat{\tilde\psi}_a(p),\hat{\tilde\psi}_b(p')]=0=
[\hat{\tilde{\psi}}^{\dag a}(p),\hat{\tilde\psi}^{\dag b}(p')],
\quad
[\hat{\tilde\psi}_a(p),\hat{\tilde\psi}^{\dag b}(p')]=
{1\over N}2\pi\delta(p-p')\delta^b_a.
\eeq	
As usual classical Poisson brackets go over to quantum commutation relations:
\beq
	\{A,B\}\to -i\hbar[\hat A,\hat B].
\eeq
In our case we will denote the parameter that measures the quantum
correction, analogous to \m{\hbar},  by \m{1\over N}.  In a minute we
will  see that this
number \m{N} must actually  be an integer. Th elimit \m{N\to \infty} is the
 classical limit.
The constraint on the observables can be implemented by restricting
attention 
to those states satisfying 
\beq
\int_0^{\infty} 
 \hat{\tilde\psi}^{*a}(p){\hat{\tilde\psi}}_a(p){dp\over 2\pi}|>=1.
\eeq

Now it is obvious that a representation for our commutation relations is 
provided by bosonic creation annihilation operators:
\beq
	[\hattilde{a}_a(p),\hattilde{a}_b(p')]=0=
	[\hattilde{a}^{\dag a}(p),\hattilde{a}^{\dag b}(p')],
\quad
[\hattilde{a}_a(p),\hattilde{a}^{\dag b}(p')]=2\pi\delta(p-q)\delta^b_a.
\eeq	
with 
\beq
	\psi_a(x)={1\over \surd N}a_a(x),\quad \psi^{\dag a}(x)=
{1\over \surd N} a^{\dag a}(x).
\eeq
Then the constraint becomes just the condition that we restrict to states 
containing \m{N} particles:
\beq
	\int_0^\infty \tilde a^{\dag a}(p)\tilde a_{p}(x){dp\over 2\pi}=N.
\eeq
This is why \m{N} must be an integer!.

Now we know that we are dealing with a system of \m{N} bosons interacting with
 each other under the hamiltonian
\beqs
	{1\over N}\hat E_1(\psi)&=&\sum_a\int_0^\infty\half[p+{\mu^2\over
p}]\tilde a^{\dag a}(p)\tilde a_{a}(p){dp\over 2\pi}\\
& & +
{\alpha_1+\alpha_2\over 2}N
\int a^{\dag a}(x)a^{\dag b}(y)a_{b}(y)a_a(x)|x-y|dxdy.
\eeqs
The classical (or large \m{N}) limit we have been discussing so far is
just the mean field approximation to this many-body problem. The
semi-classical approximation will give us the leading corrections in
the case of finite \m{N}. This basic insight is due to Witten, in a by now 
classic paper \cite{wittenlargeN}.

What are these bosons? We will see in the next chapter  that these
bosons are just the valence quarks of  the parton model, sripped of
their color!. Quarks are
of course fermions. However the wavefunction of the system must be
totally anti-symmetric in the color indices beacuse of the condition
that the state be invariant under \m{SU(N)}. Thus in the remaining
indices the wavefunction must be symmetric: if we ignore color the
valence quarks behave like bosons.

 Note that the momentum of the particles created by
\m{a^{\dag}(p)} is always positive. Thus the total momentum
\beq
\hat P=\int_0^\infty p \ a^{\dag a}(p)a_a(p){dp\over 2\pi}
\eeq
is a  positive operator. Indeed on a state containing \m{N} particles,
\beq
	|a_1,p_1;a_2,p_2;\cdots a_N,p_N>=a^{\dag a_1}(p_1)a^{\dag
a_2}(p_2)\cdots a^{\dag a_N}(p_N)|0>
\eeq
\m{P} is just the sum of individual momenta, each of which is positive:
\beq
		\hat P|a_1,p_1;a_2,p_2;\cdots a_N,p_N>=[p_1+p_2+\cdots p_N]|a_1,p_1;a_2,p_2;\cdots a_N,p_N>.
\eeq
A general state   will be, in this basis described a wavefunction
 \m{\tilde\phi}
\beq
	|\phi>=
\sum_{a_1,\cdots a_N}\int_0^\infty {dp_1\over 2\pi}\cdots {dp_N\over 2\pi}
\tilde\phi(a_1,p_1;\cdots a_N, p_N)|a_1,p_1;\cdots a_N, p_N>.
\eeq
An eigenstate of \m{\hat P} with eigenvalue \m{P} will satisfy
\beq
\left[p_1+\cdots p_N\right]\tilde\phi(a_1,p_1;\cdots a_N, p_N)=
P\tilde\phi(a_1,p_1;\cdots a_N, p_N).
\eeq
Since each of the momenta \m{p_i} are positive, it follows that they must each 
be less than the total momentum \m{P}:
\beq
	0\leq p_i\leq P,\; {\rm for} \; i=1,\cdots N.
\eeq

We will see in the next chapter 
that the \m{p_i} are the momenta of the valence partons: the proton is a 
simultaneous eignstate of \m{\hat H} and \m{\hat P}. 
We have just seen a very important point:  our  theory at finite \m{N} (but 
within the approximation of the valence parton model) predicts that the 
wavefunction must  vanish unless each of the parton momenta \m{p_i} are less 
than \m{P}. But
there is no \m{N} in this inequality; so it must hold  even in
 the large \m{N} limit!. 

We will be finding an approximate eigenstate of \m{\hat H}, by a variational
 principle:  a sort of mean field theory,  the large \m{N} limit.The naive
 choice is a  wavefunction which is a product of single particle
 wavefunctions. But such a naive version of mean field theory will violate the
 exact inequality  we just established on momentum eigenstates. We should find
 our variational approximation to the eigenstate of  the hamiltonian, within
the space of momentum eigenfunctions.

Thus we assume that the 
wavefunction is approximated by a wavefunction that is just a product
{\it except} for the constraint that the momenta add up to \m{P}:
\beq
	\tilde\psi(a_1,p_1 ;a_2,p_2;\cdots
a_N,p_N)=2\pi\delta(\sum_ip_i-P)
\tilde\psi(a_1,p_1)\tilde\psi(a_2,p_2)\cdots \tilde\psi(a_N,p_N).
\eeq
Thus the fraction of the momentum carried by each particle is less
than one:
\beq
	\psi(p)=0,\quad \;{\rm unless}\; 0\leq {p\over P}\leq 1.
\eeq
That is how we recover the fact that the quark distribution function
must vanish when the Bjorken variable is greater than one. This is a
sort of semi-classical correction  in the \m{1\over N} approximation.
We must solve the classical equations of motion for \m{\tilde\psi(p)}
subject to the boundary condition  that it vanish outside of the
interval \m{0\leq p\leq P}. This requires a modification of the
variational ansatz but it it is possible to do that.

This variant of mean field theory is rather like the micro-canonical ensemble. 
The naive mean field theory where only the expectation value of momentum is
 required to be \m{P} is like the canonical ensemble.

\section{Valence Quark Distributions}

Now we come to the point of comparison of the calculated valence
parton distribuion function againt the data. It is not in fact
necessary to make a direct comparison with the data on Deep Inelastic
Scattering. A generation of phenemenologists have extracted the
valance parton structure functions from the data. More precisely they
have assumed a parametric form such as
\beq
	\phi(x_B)=Ax_B^{\nu_1}(1-x_B)^{\nu_2}[1+a_1x_B+a_2x_B^2]
\eeq
for the parton distribution functions and fit to all known data
points. Since the data consists of measurement of the DIS (Deep
Inelastic Scattering)
cross-section at different values of \m{Q^2} this fit uses the
convolution of this distribution function with the structure functions
of the quarks calculated in perturbation theory. There are about 6
structure functions (corresponding to \m{u,\bar u,d,\bar d, s} quarks
and the and gluons ) so altogether there are about 30 parameters in
addition to the fundamental parameters of perturbative QCD. Altogether
there are about a thousand data points, coming from measurement of the
cross-section for \m{e-p}, \m{e-n}, \m{\nu-p} or \m{\nu-n} scattering
at various energies. Thus the mere extraction of these distribution
functions has itself become a subfield of particle physics
\cite{mrst,cteqhandbook,grv}. The major groups seem to be in
general agreement with each other, although a detailed 
analyzis of  the errors in their parameters is not yet available. 
There are recent attempts to
estimate the systematic and statistical errors, but they seem incomplete.

The comparison should be made with the weighted average over the \m{u} and
\m{d} quark distributions,weighted so as to get an isospin invariant
combination. This is because we have not yet done a collective
variable variable quantization of the isospin degrees of freedom of
the proton, so  the distribution function we are computing is  the isospin
invariant one.

Another complication in the comparison is that not all the momentum in
the baryon is carried by the valence quarks: it is known that only
about half of the momentum is in the valence quarks, the rest being in
the anti-quarks (or `sea' quarks) or gluons. This affects our momentum
sum rule. Essentially what it does is that the \m{N} in the sumrule
gets replaced by \m{N_{\rm eff}={N\over f}} where \m{f} is the fraction
of the momentum carried by the valence quarks. Until we do a complete
calculation allowing for the contributions of the sea quarks and the
gluons (i.e., without assuming the factorizable ansatz for \m{M} and
without ignoring the scalar fields \m{\phi_A}) we must treat \m{f} as
a parameter and choose the value that gives the best fit. The good
news is that it is the only  parameter:\m{m^2} is fixed to be zero\footnote{This by the way
means that \m{\mu^2=m^2-{\tilde g^2\over \pi}} is negative. There is no
contradiction here, since the quark is not a true particle: the mass
of the baryon is still predicted to be real.},
since the up and down quarks are known to have masses that are smal
compared to the QCD scale \m{\alpha_1}. (\m{m_u\sim 5\; MeV} and
\m{m_d\sim 10\; Mev} while \m{\alpha_1\sim 100\; MeV}.) 

The only other parameter of the theory is \m{\tilde g^2} ; but since the
distribution functions are dimensionless function of a dimensionless
variable, the dependence on \m{\alpha_1} cancels out.
 
In comparison with experimental data we should keep in mind that we
are in fact ignoring some of the constituents of the baryon: the
valence quarks are knwon to carry only about half of the momentum of
the baryon. The rest is in the sea quarks and the gluons. Thus we
introduce a parameter \m{f}, the fraction of the momentum carried by
the valence quarks. We find that for a value of \m{f=0.6} the
wavenfunction predicted by us agrees quite well with experiment. (The
value of being about a half is consistent with other ways of looking
at the situation.) Thus we have solved the problem of deriving the quark 
structure functions of the baryon from QCD.

\begin{figure}
\centerline{\includegraphics{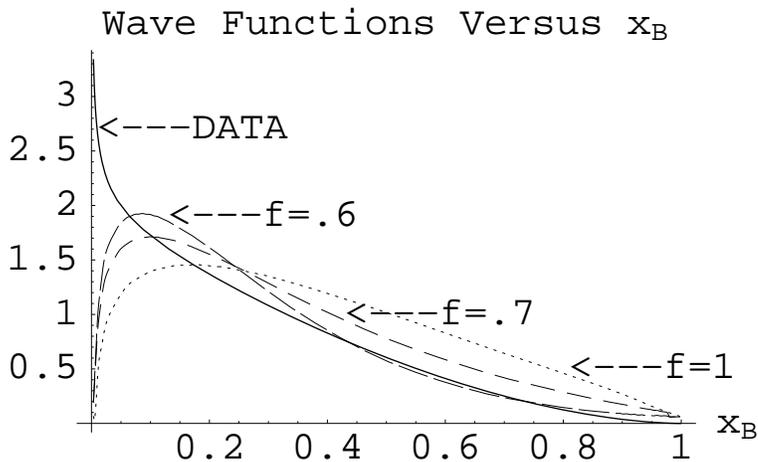}}
\caption{Comparison of  parton wavefunctions \m{\surd\phi(x)}. The MRST
 distribution agrees with our prediction best when the fraction \m{f} 
of the momentum carried by the valence quarks is about \m{0.6} .}
\end{figure}

It still remains to calculate the anti-quark and gluon structure functions. 
We will turn to that in later publications. In a later  chapter we will give some ideas that help in solving this problem.

In the next chapter we will see how to reconcile the soliton model with the 
parton model.

\chapter{The  Parton Model}

This chapter is based on the paper \cite{krishraj} 
which in turn was an expansion of some short  statements  made in 
Ref. \cite{2dhadron}.

We have been describing the idea that the baryon is a soliton,
essentially made up of an infinite number of mesons. But this seems to
contradict the simple and successful picture of the baryon as a bound
state of quarks. At the time the Skyrme model was revived, it was shown that
the static (low energy properties) of the baryon that are so well
explained by the quark model (magnetic moments, mass differences,
flavor multiplet structure) can all be rederived in the soliton
model\cite{baltasi}. But the only explanation for the structure of the
proton in deep inelastic scattering still appeared to be the
quark-parton model. In these lectures, we are describing how to
explain the structure functions within the soliton model as well. In
fact we will be able to go beyond the simple minded parton model.

It is an old idea in the study of   the structure of the proton that the
constituents-the partons- can be thought of as free particles. This
then raises the question of how they bind to form the proton in the
first place. It is the binding mechanism that determines the
wavefunction of the partons and hence the structure functions of the
hadron. In this section we will present a model of interacting
partons. Perhaps surprsingly, it will turn out to be equivalent to the
soliton model in the factorizable ansatz (and in the large \m{N}
limit). Thus the parton model is merely an approximation to the
soliton model. Later, by a deeper analysis of the soliton model, we will
derive  the sea quark and gluon distributions as well.

It is at the moment impossible to derive the particle spectrum of four
dimensional QCD directly. There are many attempts to do compute the spectrum by
direct numerical simulation of QCD, but they have to yet to surmount
many obstructions. For example, the asymptotic freedom of QCD implies
that the energy of all physical states are exponentially small
compared to the cut-off. Thus extra-ordinary accuracy is required of
all direct numerical computations of the energy, a problem that is
only exacerbated by the large ( in principle infinite) number of
degrees of  freedom in the system. However, it may be possible to
understand the spectrum of two dimensional QCD by diagonalizing its
hamiltonian numerically. Being a finite (rather than asymptotically
free) theory  numerical calculations are much more reliable.

\section{Parton Wavefunctions}

We will follow largely the discussion in \cite{krishraj}
The valence parton model assumes that the proton is made of \m{N}
partons (quarks) which are fermions transforming under the fundamental
representation of color. The idea of color of course was not present
in the original version of this model, due to Bjorken and
Feynman. Also it was not known then that the partons were spin half
particles and hence fermions. But these were straightened out soon
with the identification of the valence partons with the quarks. \
footnote{It is 
known that  only about half of the momentum of the baryon is accounted for by
these valence partons: the rest must be carried by the `sea quarks'
 and by the gluons.}

We will work think of the parton wavefunctions as functions of the
null momentum \m{p=p_0+p_1}. In addition, the partons  carries the
quantum numbers of flavor and spin (together denoted by \m{a}) and
color \m{\alpha}. Thus the wavefunction of a single parton will be
\m{\tilde\psi(a,\alpha,p)}. Since \m{p\geq 0} for the null component
of momomentum, we require that \m{\tilde\psi} vanish for negative
\m{p}.

 A baryon is made of \m{N} such partons so its wavefunction is a
completely antisymmetric function
\m{\tilde\psi(a_1,\alpha_1,p_1;a_2,\alpha_2,p_2;\cdots;
a_N,\alpha_N,p_N)}.
However, the baryon must be invariant under color: transform under the
trivial representation of color \m{SU(N)}. This means that the
wavefunction is completely antisymmetric in color alone: 
\beq
	\tilde\psi(a_1,\alpha_1,p_1;a_2,\alpha_2,p_2;\cdots,
 a_N,\alpha_N,p_N)=\eps_{\alpha_1,\alpha_2,\cdots \alpha_N}
\tilde\psi(\alpha_1,p_1;\alpha_2,p_2;\cdots \alpha_N,p_N)
\eeq
In other words, the wavefunction is completely {\it symmetric} in
spin, flavor and momentum quantum numbers. If we factor out color from
the wavefunction, the partons behave as if they are {\it bosons}.

\section{Hamiltonian}

The kinetic energy of a single parton is
\beq
	\half\big[p+{\mu^2\over p}\big].
\eeq
Hence the kinetic energy of the partons making up a baryon is
\beq
\sum_{a_1\cdots a_N}\int_0^\infty \sum_{i=1}^{N}\half
[p_i+{m_{a_i}^2\over
p_i}]|\tilde\psi(a_1,p_1;\cdots; a_N,p_N)|^2
{dp_1\cdots dp_N\over (2\pi)^N}.
\eeq

In the early versions of the quark-parton model, they were treated as
if they are free particles. That cannot be quite correct, since they
must after all bind to form the baryon. The simplest idea would be to
let them interact through a pairwise potential \m{\alpha_1v(x)}:
\beq
\half \alpha_1\sum_{a_1\cdots a_N}\int_0^\infty 
\sum_{i\neq j }
v(x_i-x_j)|\psi(a_1,x_1;\cdots; a_N,x_N)|^2 dx_1\cdots
dx_N.
\eeq

Now we must allow for a self-energy term as well in the
hamiltonian. The point is that the effective mass of the partons may
not be the same as their bare mass. Just as the electrons in a metal
have a different effective mass from the free electron, the effective
mass of the parton inside a baryon may be different from that of a
`bare' parton: there can be a finite renormalization of the mass.
 We can allow for this by replacing the \m{\mu^2} in the
kinetic energy by an effective mass \m{\mu^2}. 

Since the partons
(quarks) cannot be separated out to infinity, one could wonder what
the `bare' parton mass \m{m} means: it has no meaning as the mass of
any particle that can exist in an asymptotic state. However, it is
possible to give the bare parton mass a meaning in terms of high
energy processes that do not involve asymptotic states.
 For example the weak interaction can be used as
a probe of the quark masses: the weak decay rates of the quarks are
sensitive to the bare masses. This gives a way to make sense of the
bare parameter \m{m}. But for our purposes, what matters is the
effective mass \m{\mu}.

Thus the hamiltonian of the valence quark model is
\beqs
	{\cal E}_N(\tilde\psi)&=&\sum_{a_1\cdots a_N}\int_0^\infty \sum_{i=1}^{N}\half[p_i+{\mu_{a_i}^2\over
p_i}]|\tilde\psi(a_1,p_1;\cdots a_N,p_N)|^2{dp_1\cdots
dp_N\over (2\pi)^N}\cr
 & &+ \half \alpha_1 \sum_{a_1\cdots a_N}\int_0^\infty
\sum_{i\neq j}
v(x_i-x_j)|\psi(a_1,x_1;\cdots a_N,x_N)|^2 dx_1\cdots
dx_N.\nonumber
\eeqs
The ground state of this hamiltonian is the baryon of the valence
quark model. 

We havent yet decided what potential \m{v(x)} to use. There is ample
evidence that the quark-quark potential is linear \cite{potmodel}. In
any case as in QHD, Lorentz invariance will lead to this choice n our
lightcone co-ordinates. So we choose \m{ v(x)=\half |x|}.

With this choice we get  exactly the same hamiltonian  we had in the last
 chapter, from 
the quantization of the rank one ansatz with \m{1\over N} playing 
the role of \m{\hbar }. Thus  the rank one approximation to 
 the topological soliton model is equivalent to this interacting valence
 parton model.

\section{Hartree Ansatz}

The ground state of a many boson system can often be described by mean
field theory: each boson moves in the field created by all the
others. Moreover, all the bosons can be assumed to occupy the same
single particle state in this ground state. After the color is
factored out, the partons in our model behave just like bosons. Hence
we should be able to simplify the problem by making this mean field
approximation. More specifically,
\beq
	\tilde\psi(a_1,p_1;\cdots a_N,p_N)=
2\pi\delta(\sum_ip_i-P)\prod_{i=1}^N\tilde\psi(a_i,p_i).
\eeq
Here \m{P} is the total momentum of the system.
Since \m{p\geq 0} we must impose
\beq
	\tilde\psi(p)=0
\eeq 
for \m{p\leq 0}. It then follows that \m{p} is less than the total
momentum \m{P}.

The wavefunction satisfies the normalization condition
\beq
	\int_0^P |\tilde\psi(p)|^2{dp\over 2\pi}=1
\eeq
and the momentum sum rule:
\beq
	N\int_0^P p|\tilde\psi(p)|^2{dp\over 2\pi}=P.
\eeq

The energy is calculated by putting this ansatz into the earlier formula:
\beqs
	E&=&\sum_a\int_0^\infty \half[p+{\mu_a^2\over
p}]|\tilde\psi(a,p)|^2{dp\over 2\pi}+\cr
  & & \half{\tilde{g}^2}\int_{-\infty}^\infty v(x-y)
\sum_a|\psi(a,x)|^2\sum_b|\psi(b,y)|^2dxdy.
\eeqs

At this point we notice that we have {\it exactly} the same problem as 
in the last chapter. The energy of the rank one  ansatz for
\m{\tilde M(p,q)} is exactly the same as that of the parton model in
the Hartree approximation. Thus the mean field theory of the valence 
 parton model is our {\it classical} theory on the projective space
 \m{CP({\cal H}_+)}.. This interpretation of the
rank one  ansatz in terms of the parton model was already noted in
Ref. \cite{2dhadron}.

What would be the leading  deviation  away from the large \m{N} or classical
 limit? We argued in the last chapter in exactly the same theory 
(but viewed as  the quantization of the rank one ansatz) that it amounts to the
ansatz
\beq
\tilde\psi(p_1,\cdots p_N)=
2\pi\delta(p_1+p_2\cdots p_N-P)\prod_i\tilde\psi(p_i).
\eeq
There are now some correlations between the quarks: each momentum is less than 
\m{P} and also \m{p_N} can be eliminated in terms of the others. The latter is
 a small effect; but the inequalities on the momenta, \m{0\leq p_i\leq P} 
cannot be ignored.

\chapter{ Beyond the  Valence Parton Model }

We saw that the separable ansatz in the soliton theory corresponds to
the valence parton approximation. Now we see how to go beyond that and
get the true minimum of the energy. We can  improve on the valence 
 parton model by adding Sea quarks and anti-quarks; this requires us to 
invent an improved variational ansatz. In another direction we can minimize 
 the energy in the Grassmannian by numerical methods. We will describe ideas in both directions. 

\section{ A Co-ordinate System }
Recall that the equations to be solved are
\beq
	[E'(M),\eps+M]=0,
\eeq
along with the constraints:
\beq
	[\eps+M]^2=1,\quad \tr_\eps M =-2.
\eeq

Define 
\beq
	\eps_1=\eps-2\psi\otimes\psi^{\dag},
\eeq 
where \m{\psi} is the minimum in the subset of separable
kernels; or at least a good approximation to it.
 Rather than parametrize the soliton by 
 the deviation from the vacuum \m{\eps} it
makes  more sense to use the deviation from our approximate solution
\m{\eps_1} which is in the same connected component. 
So we define a new variable \m{M_1} by 
\beq
	\eps+M=\eps_1+M_1.
\eeq
Then the constraints on \m{M_1} are 
\beq
	[\eps_1,M_1]_+ +M_1^2=0,\quad \tr_\eps M_1=0.
\eeq
The trace of \m{M_1} is zero since the baryon number is  now already
carried by \m{\eps_1}: 
\beq
	-\half \tr[\eps_1-\eps]= 1.
\eeq
Let \m{W} be the subspace where \m{\eps_1} takes eigenvalue \m{-1}:
\beq
	W={\cal H}_-\oplus <\psi>
\eeq
where \m{<\psi>} denotes the linear span of \m{\psi}. We have an
orthogonal splitting of \m{\cal H} according to this splitting.
We can now represent \m{M_1} in terms of a co-ordinate system centered at the point \m{\eps_1}:
\beq
	M_1=-2\pmatrix{[1+ZZ^{\dag}]^{-1}-1&[1+ZZ^{\dag}]^{-1}Z\cr
		     Z^{\dag}[1+ZZ^{\dag}]^{-1}&Z^{\dag}[1+ZZ^{\dag}]^{-1}Z}
\eeq
Here, \m{Z:W^\perp\to W} is an arbitrary Hilbert-Schmidt operator. It
is straighforward to check that this is a solution to the constraint
equations. Moreover, in some finite neighborhood of \m{\eps_1}, all
points of the Grassmannian can be represented this way. Of course this
co-ordinate system will break down if we go too far away from
\m{\eps_1}: we are assuming that the true minimum lies close enough to
the approximate solution to be in this co-ordinate chart.

Given a fixed \m{\psi}, the operator \m{Z} above provides a co-ordinate system on an neighborhood of the Grassmannian.
This co-ordinate system 
 only covers a part of the connected component with
 baryon number one. It takes a countably infinite number of  such charts 
(corresponding to different choices of \m{\psi}) to cover the whole space of
 baryon number one configurations.

\section{The Method  of Steepest Descent}

We need to minimize the function \m{E(M)} subject to the constraints
\beq
	[\eps,M]_++M^2=0,\quad -\half \tr M=1,\quad \tr M^2<\infty.
\eeq
A simple method for minimizing functions of several variables is
steepest descent: we start at some initial point  and move along the
straight line 
 opposite to the gradient of the function at that point a small
distance. Then we recompute the gradient at the new point and repeat
the procedure. If the topography of the constant energy surfaces is
not too  complicated we will eventualy arrive at the minimum. 

The main complication in our case is of course the constraint: the
gradient vector \m{E'(M)} is not tangential to the Grassmannian so if
we move opposite to it we will leave the constraint surface. Even if
we project out the tangential component of the vector, we will still
leave the surface if we move along the straightline in that
direction. The proper geometric solution to this problem is to move a
small distance 
along the geodesic at \m{M} tangential to the gradient vector. Then we
will recompute the gradient and find the geodesic at the new
point. This is feasible beacuse the Grassmannian is a homogenous
manifold and we can find the geodesic on it easily using its high
degree of symmetry.

More explicitly, recall that the tangential projection of the gradient
vector is 
\beq
	T={1\over 4}[\eps+M,[\eps+M,E'(M)]].
\eeq
The vector 
\beq
	Y={1\over 2}[\eps+M,E'(M)]
\eeq
is at right angles to the tangential part of the gradient: it is in
fact obtained by multplying \m{T} by the complex structure of the
Grassmannian. Just as in the case of the sphere, the geodesic is
obtained by rotating the point about an axis orthogonal to the tangent vector. 
More explicitly,
the geodesic starting at \m{M} tangential to \m{T} is
\beq
	\gamma(\tau)=e^{\tau Y}[\eps+M]e^{-\tau Y}-\eps.
\eeq
Thus the steepest descent algorithm is

\noindent
1. Choose an initial configuration \m{M_0} and small parameter \m{\tau}.

\noindent
2. Given the \m{k^{\rm th}} configuration \m{M_k}, calculate the
   gradient \m{E'(M_k)} and \m{Y_k=\half [\eps+M_k,E'(M_k)]}.

\noindent
3. Set \m{M_{k+1}=e^{\tau Y_k}[\eps+M_k]e^{-\tau Y_k}-\eps} and repeat
   the previous step.

The value of \m{\tau} has to be chosen by some trial and error. Too
small a value will produce changes in the configuration within the
noise due to numerical errors. Too large a choice will not give a
convergent sequence: we will bounce around all over the
Grassmannian. But we found that  in practice,  a proper value of \m{\tau} 
 can be found quickly.

How do we choose the initial configuration \m{M_0}? Since we believe
that the valence approximation is good, we could use as the starting
point a separable configuration, minimizing the energy within that subspace.

\section{  Sea Quarks and Anti-quarks}

We saw that if we restrict the dynamics of our theory to the rank one ansatz,
\m{M=-2\psi\otimes \psi^{\dag}}, we get the valence parton model.
To get a more general picture that includes Sea quarks and anti-quarks 
(but is still not the total picture) we must use the ansatz with rank three.
 In fact by going to ansatzes of larger and larger rank we can get better and
 better approximations to the baryon structure functions.

The rank three ansatz is
\beqs
	-\half M&=&\psi_0\otimes \psi_0^\dag +
\zeta_-\big\{\zeta_-[\psi_+\otimes\psi_+^\dag -\psi_-\otimes\psi_-^\dag]\cr
 & & -
\surd[1-\zeta_-^2][\psi_-\otimes\psi_+^\dag +\psi_+\otimes\psi_-^\dag]\big\}
\eeqs
where \m{\psi_-,\psi_0,\psi_+} are three vectors in \m{\cal H} satisfying
\beq
	\eps\psi_-=-\psi_-,\quad  \eps\psi_0=\psi_0,\quad 
\eps\psi_+=\psi_+,
\eeq
\beq
	||\psi_-||^2=||\psi_0||^2=||\psi_+||^2=1,\quad \psi_0^\dag\psi_+=0.
\eeq       
Moreover,
\beq
	0\leq \zeta_-^2\leq 1.
\eeq
We saw that \m{\zeta_-^2} is the probability of finding an anti-quark inside the
baryon. 

As noted earlier, the anti-quark distribution function is just 
\m{\zeta_-^2|\tilde\psi_a(-p)|^2}.
The baryon number  is
\beq
	B=\sum_a\int_0^\infty\big\{|\tilde\psi_{0a}(p)|^2+
\zeta_-^2[|\tilde\psi_{+a}(p)|^2-\tilde\psi_{-a}(-p)|^2]\big\}{dp\over 2\pi}.
\eeq
The total momentum is, similarly,
\beq
	P=\sum_a\int_0^\infty p\big\{|\tilde\psi_{0a}(p)|^2|
+\zeta_-^2[|\tilde\psi_{-a}(-p)|^2+|\tilde\psi_{+a}(p)|^2]\big\}{dp\over 2\pi}
\eeq
These confirm the interpretation of \m{\psi_0} as the valence quark wavefunction and \m{\psi_+(p)} as the Sea quark wavefunction.	

Also,  the kinetic energy  is
\beq
	K=\sum_a\int_0^\infty \half[p+{\mu^2\over p}] \big\{|\tilde\psi_{0a}(p)|^2|
+\zeta_-^2[|\tilde\psi_{-a}(-p)|^2+|\tilde\psi_{+a}(p)|^2]\big\}{dp\over 2\pi}
\eeq

The potential energy is more complicated. Recall that the potential
energy is simpler in position space while the kinetic energy is
simpler in momentum space. Our wavefunctions now depend on a discrete
variable in addition to momentum, and there is a unitary
transformation in these discrete variables that is the counterpart to
Fourier transformation. 
This is the transformation to  a basis in
which \m{\mu} is diagonal; the wavefunctions will no
longer be eigenstates of \m{\eps}.

\beq
	-\half M=\psi_0\otimes \psi_0^\dag +
\zeta_-\big\{\psi_1\otimes\psi_1^\dag-\psi_2\otimes\psi_2^\dag\big\}
\eeq
where
\beq
\psi_1={1\over \surd2}\bigg\{\surd[1-\zeta_-]\psi_--\surd[1+\zeta_-]\psi_+\bigg\}
\eeq
\beq
\psi_2={1\over \surd 2}\bigg\{\surd[1+\zeta_-]\psi_-+\surd[1-\zeta_-]\psi_+
\bigg\}
\eeq

The potential energy is then,
\beqs
U&=&{\alpha_1\over 2}\sum_a\int \bigg\{|\psi_{0a}(x)|^2 V_{00}(x)+\cr
& & \zeta_-^2|\psi_{1a}(x)|^2 V_{11}(x)+\zeta_-^2|\psi_{2a}(x)|^2 V_{22}(x)\bigg\}dx\cr
& & +{\alpha_1\over 2}2\;{\rm Re}\; \sum_a\int   \bigg\{\zeta_-
\psi_{0a}(x)\psi_{1a}^*(x) V_{10}(x)-\cr
& & \zeta_-\psi_{0a}(x)\psi_{2a}^*(x) V_{20}(x)-\zeta_-^2\psi_{1a}(x)\psi_{2a}^*(x) V_{21}(x)\bigg\}dx.
\eeqs
The mean fields are determined by solving the differential equations
\beq
V_{\alpha\beta}''(x)=\psi_{\alpha a}(x)\psi_{\beta a}^*(x).
\eeq
with the boundary conditions
\beq
V_{\alpha\beta}(x)\to \delta_{\alpha\beta}\half|x|,\;{\rm for}\; |x|\to \infty.
\eeq
Equivalently, 
\beq
	V_{\alpha\alpha}(0)=V_{\alpha\alpha}'(0)=0
\eeq
for the diagonal components, and,
\beq
	V_{\alpha\beta}(\infty)=V_{\alpha\beta}'(\infty)=0\;{\rm for}\; 
\alpha\neq \beta
\eeq	
for the off diagonal components.

By choosing appropriate variational ansatzes we can estimate the
anti-quark content of the proton. Or we can derive integral equations
for the functions \m{\psi_{\pm,0}} and solve them numerically. Both
methods are being pursued. Initial results are encouraging. Detailed
results will appear in a separate publication.

\centerline{\bf Acknowledgement}

I thank the students and postdoctoral fellows at Rochester who have worked 
withme at various stages in implementing this research program: 
P. Bedaque, K. S. Gupta, S. Guruswamy, I. Horvath,V. John,  G. Krishnaswami, 
C-W. H. Lee, O. T. Turgut. I have also benefitted from discussions with 
A. P. Balachandran, A. Bodek, J. Mickelsson, S. Okubo, D. Pickrell  and 
J. Schechter. I am especially grateful to O. T. Turgut and V. John for 
proof-reading an early version of these notes. Also I thank C. Saclioglu and 
J. Mickelsson for invitations to Istanbul and Stockholm and making my visits
 there so enjoyable. This work is supported by the Department 
of Energy.

\appendix

\chapter{ A Null  Co-ordinate System}

\section{Kinematics}

It will be convenient to use a co-ordinate system that combines the
advantages of the null and Cartesian co-ordinate systems.

If \m{x^0} and \m{x^1} are the usual Cartesian co-ordinates in
Minkowski space, the metric is
\beq
	ds^2=\big[dx^0\big]^2-\big[dx^1\big]^2.
\eeq
We define now
\beq
	t=x^0-x^1,\quad x=x^1
\eeq
so that 
\beq
	ds^2=dt[dt+2dx].
\eeq
The Minkowski metric is,
\beq
	\eta_{\mu\nu}=\pmatrix{1&1\cr
		      1&0}
\eeq
Thus the vector \m{{\pdr\over \pdr t}} is time-like while  
\m{{\pdr\over \pdr x}} is null. The initial values of fields will be
given on a  surface of  constant \m{t}, which is a null line.

Momentum \m{p=p_\mu dx^\mu=p_0dt+pdx} is a \m{1}-form (co-vector); we
will use the same letter to denote the momentum 1-form as well as its
null componenet, but it should be clear from the context which one we
mean. To find its magnitude we
must use the inverse of the above metric tensor:
\beq
	\eta^{\mu\nu}=\pmatrix{0&1\cr
		      1&-1},\quad p^2=p_\mu p_\nu\eta^{\mu\nu}=2p_0p-p^2.
\eeq
Thus the mass shell condition becomes
\beq
 p_0=\half\bigg[p+{\mu^2\over p}\bigg].
\eeq
Here, \m{\mu} is the rest mass of the particle.

We see now one of the main technical advantages of using the
null-time co-ordinate system: energy and momentum have the same
sign. In the usual space-time co-ordinates,
\m{p_o=\pm\surd[p_1^2+\mu^2]} and therefore no such simple
relationship exists. In the Dirac theory of fermions the states of
negative energy are  occupied; this becomes merely the condtition that
the negative momentum states be occupied. 

\section{Dirac Matrices}

The Dirac matrices are best thought of as matrix-valued vectors, since
they appear in the form \m{\gamma^\mu\nabla_\mu} in the action. Thus
the Dirac algebra
\m{\gamma^\mu\gamma^\nu+\gamma^\nu\gamma^\mu=2\eta^{\mu\nu}}
  becomes
\beq
	\big[\gamma^t\big]^2=0,\quad
\gamma^t\gamma+\gamma\gamma^t=2,\quad \gamma^2=-1.
\eeq
We will choose the explicit representation
\beq
	\gamma^t=\pmatrix{0&2\cr
			  0&0},\quad \gamma=\pmatrix{0&-1\cr
			  1&0}.
\eeq
In any representation of Dirac matrices there is a `charge-conjugation'
matrix \m{C} satisfying
\beq
	C\gamma^\mu C^{-1}=\left(\gamma^{\mu}\right)^{T}.
\eeq
In the usual space-time formalism this matrix is often \m{\gamma^0}
itself, but that is a representation-dependent fact. In our
representation,
\beq
	C=\pmatrix{0&1\cr
			  1&0}.
\eeq
 Given the  Dirac spinor \m{q=\pmatrix{q_1\cr 
					 q_2}}, the conjugate spinor is
\beq
	\bar q=q^{\dag}C=\pmatrix{q_2^\dag & q_1^\dag}.
\eeq

\section{Free Fermions}

The Lagrangian  of a free Dirac fermion becomes
\beqs
	L_D&=&\bar q\gamma^\mu[-i\pdr_\mu]q+m\bar qq\cr
	   & &+ 
2q_2^\dag[-i\pdr_t]q_2+q_1^\dag(-i\pdr_x)q_1-q_2^\dag(-i\pdr_x)q_2+
m[q_2^\dag q_1+q_1^\dag q_2].
\eeqs
We see that the field \m{q_1} has no dynamical degrees of freedom: it
has no time derivative in the action. Hence it can be eliminated by
its equation of motion:
\beq
	q_1=-{m\over \hat p}q_2
\eeq
where \m{\hat p=-i\pdr_x}. Putting this  back 
into the action
and changing variables
\beq
	\chi= \surd 2 \; q_2
\eeq
 gives us
the effective action for the propagating field:
\beq
	L_\chi=\chi^\dag(-i\pdr_t)\chi-\chi^{\dag}\half\big[\hat
p+{m^2\over \hat p}\big]\chi.
\eeq

\section{Gauge Fields}

The lagrangian of two dimensional QCD is 
\beqs 
	L&=& {N\over {4\alpha_1}}\int \tr F_{\mu\nu}F^{\mu\nu}+
\sum_{a=1}^{N_f}\int \bar
q^{a\alpha}[-i\gamma\cdot\nabla+m_a]q_{a\alpha}.
\eeqs
 We have added in the flavor indices \m{a,b} and the color indices \m{i,j}.

 The freedom of gauge transformations can be utilized partially to impose
 the null gauge condition
\m{A_x=0}.
 Then, the fermionic part of the lagrangian becomes
\beqs
	L_D&=&
	\bar q^{ai}\gamma^\mu[-i(\pdr_\mu\delta^j_i+A_{\mu i}^j)]q_{aj}+
		m\bar q^{ai}q_{ai}\cr
	   &+& 
\chi^{\dag ai}[-i(\pdr_t+A_{ti}^j)]\chi_j\cr
& & +q_1^{\dag ai}(-i\pdr_x)q_{1ai}-\half\chi^{\dag ai}(-i\pdr_x)\chi_{ai}+
{m\over \surd 2}[\chi^{\dag a i} q_{1 ai}+q_1^{\dag a i}\chi_{ai}].
\eeqs
which becomes
\beq
	L_\chi=\chi^{\dag a}(-i\pdr_t)\chi_a-\chi^{\dag a}\bigg\{\half\big[\hat
p+{m^2\over \hat p}\big]-iA_t\bigg\}\chi_a
\eeq
upon eliminating \m{q_1}.

To this we must add the 
  action of the Yang--Mills field itself, which  looks quite simple in
 this co-ordinate
system and gauge:
\beq
	L_{YM}={N\over 4\alpha_1}\tr
  F_{\mu\nu}F_{\rho\sigma}\eta^{\mu\rho}\eta^{\nu\sigma}={N\over 2\alpha_1}\tr[\pdr_xA_t]^2.
\eeq
The field \m{A_t} does not propagate and can be eliminated.
Thus the action of two-dimensional QCD can be written entirely in terms
of the field \m{\chi}. 

\section{The Dirac Vacuum}

Let us return to the free fermion theory with Lagrangian
\beq
	L=\chi^{\dag}(-i\pdr_t)\chi-\chi\half\big[\hat
p+{m^2\over \hat p}\big]\chi
\eeq
Upon quantization, the field \m{\chi} becomes an operator satisfying
the fermionic anti-commutation relations
\beq
	[\chi(x),\chi(y)]_+=\delta(x-y),\quad 
[\chi(x),\chi(y)]_+=0.
\eeq

If there were only a finite number of operators, such canonical
 anti-commutation relations would have a unique representation. In the
 infinite dimensional case physical ideas have to be brought in to
 choose the right representation. Dirac showed that the correct choice
 is to assume that all the negative energy states are  filled even in the
 vacuum. Since energy and momentum have the same sign in our
 co-ordinate  system, this condition is easy to implement. We define
 the  vacuum by 
\beq
\tilde\chi^{\dag}(p)|0>=0\;{\rm for}\; p<0,\quad 
\tilde\chi(p)|0>=0\;{\rm for}\; p>0.
\eeq
We have defined the Fourier transforms
\beq
	\chi(x)=\int \tilde\chi(p)e^{ipx}{dp\over 2\pi}
\eeq
etc.
Then the normal ordered product of a pair of operators is defined as 
\beq
	:\chi^{\dag}(p)\chi(p'):\; =\chi^{\dag}(p)\chi(p')
\eeq
unless both \m{p} and \m{p'} are negative, in which case it is 
\beq
:\chi^\dag(p)\chi(p'):\; =-\chi(p')\chi^\dag(p).	
\eeq
The point is that then the expectation value of 
normal ordered current operators are zero in the 
Dirac vacuum. Indeed,
\beqs
<0|\chi^{\dag}(x)\chi(y)|0>&=&\int {dp\over 2\pi}{dq\over 2\pi}e^{-ipx+iqy}<0|\tilde\chi^{\dag}(p)\tilde\chi(q)|0>\cr
&=& \int_{-\infty}^0{dp\over 2\pi}e^{ip(y-x)}=\half[\delta(x-y)+\eps(x-y)].
\eeqs 
Here, 
\beq
	\eps(x-y)=\int \sgn(p)e^{ip(x-y)}{dp\over 2\pi}={1\over \pi i}{\cal P}{1\over x-y}
\eeq
is  to be thought of as  a distribution. It is (upto a  factor of \m{i}) 
the kernel of  a well-known integral transform, the Hilbert transform.

Thus we should regard the current operators of the fermionic theory as defined
 with the normal ordering. For example, the equation of motion of the \m{A_t}  will be, in the quantum theory,
\beq
	-\pdr_x^2A_{t j}^i(x)={1\over N}\alpha_1
:\chi^{\dag a i}(x)\chi_{ a j}(x):.
\eeq
We can use this to eliminate the gauge field from the theory
 completely. Thus  two dimensional QCD can be written as a theory of
 fermions interacting with each other through a Coulomb-like
 potential. The further analysis of this theory is carried out in the
 text, towards the end of the first chapter.

\chapter{ Operator Ideals}

Here we give the basic definitions of the operator ideals we use in the text.
 A deeper discussion may be found in Ref. \cite{presseg, 
GohbergKrein, Simon}; we
 give only a rough  outline of the theory. It  is the author's fervent 
hope that  experts in functional analysis will {\it not} read this appendix.

\section{Compact and Hilbert-Schimdt Operators}

The {\it rank} of an operator \m{A:{\cal H}\to {\cal H}} on a complex
Hilbert space is the dimension of its range; i.e., the dimension of
the subspace of all vectors that can be written as \m{Au} for some
\m{u\in {\cal H}}. When the rank of \m{A} is finite, it can be thought
of as a sort of `rectangular matrix' with (possibly) an infinite
number of columns but only a finite number of rows; at least there is
a basis in which it has this form. 

For such an operator, we can define several measures of its size
(norm). For example (the operator norm),
\beq
	|A|=\sup_{u}{||Au||\over ||u||}.
\eeq
Another (the Hilber-Schmidt or H-S norm) is 
\beq
	|A|_2=\tr [AA^{\dag}]^{1\over 2}.
\eeq
\m{|A|_2^2} is also the sum  of the absolute squares of all the matrix
 elements in any basis.

The completion of the space of finite rank operators in the operator norm is
 the space of {\it compact } operators. In other words, a compact operator is
 one that can be approximated arbitrarily closely by finite rank operators,
  distance between operators being  measured with the operator norm.
If we instead complete in the H-S  norm, we get the space of Hilbert-Schimdt
 operators. 

A bounded operator is one which has finite operator norm; i.e., 
\beq
	|A|=\sup_{u\in {\cal H}}{||Au||\over |u|}
\eeq
exists. The set of bounded operators on \m{\cal H} is an algebra \m{{\cal B}({\cal H})}. 
Not all bounded operators are compact; for example the identity is
 bounded yet not compact.

A compact operator has the expansion, with \m{\mu_n>0},
\beq
	A=\sum_{n=1} \mu_n |\psi_n><\phi_n|
\eeq
the sum being either finite (when \m{A} is of finite rank ) or 
infinite (more generally). The numbers \m{\mu_n} are called singular values; if \m{A} is positive they are its eigenvalues. \m{A} and its adjoint 
\m{A^{\dag}} have the same singular values.

 Roughly speaking, the operator norm of
 \m{A} is the largest  of its singular values, while the
 Hilbert-Schmidt norm is the sum of the  
squares of the singular  values. The singular values of a compact operator
 form a sequence that converges to zero; in fact zero is the only limit point
 of the sequence. Thus, if the H-S norm  is finite, the eigenvalues must be  
tend to zero. 
In fact all H-S operators are compact.

\section{Schatten Ideals}

Many other norms can be defined in terms of the singular  values. 
For example the trace class operators are those for which the sum of 
singular values is convergent. This is stronger than the requirement that the
 diagonal elements in some basis form a summable sequence: being
trace-class  reqires  a sort  of absolute convergence. The product of
two  Hilbert-Schimdt operators 
in trace-class.

More generally, for \m{p\geq 1}, 
 we have the class \m{{\cal I}_p({\cal H})} of operators for 
which the sum \m{\sum\mu_n^p} converges. Of course \m{{\cal I}_2} is the space
 of H-S operators and \m{{\cal I}_1} that of trace class operators. The space 
of compact operators can be thought of as the limiting case
 \m{{\cal I}_\infty} and that of finite rank rank operators as the opposite limit, \m{{\cal I}_0}. We have the inclusions
\beq
{\cal I}_0\subset {\cal I}_1\subset {\cal I}_2\cdots {\cal I}_\infty\subset 
{\cal B}.
\eeq

It is very important for us that the  \m{{\cal I}_p}  are {\it two-sided 
ideals} in the algebra  of bounded 
operators; i.e., that \m{A\in {\cal B}} and \m{B\in {\cal I}_p} implies that 
both \m{AB\in {\cal I}_p} and \m{BA\in {\cal I}_p}.
These are called the Schatten ideals. We are especially interested,
 of course, in the cases \m{p=1,2}.

\section{The  Restricted Unitary Group and its Grassmannian}

In the text we are interested in the case of a Hilbert space \m{{\cal H}} 
with a given orthogonal splitting into two infinite dimensional orthogonal 
subspaces: \m{{\cal H}={\cal H}_-\oplus {\cal H}_+}. Recall that the operator
 \m{\eps} is defined to have eigenvalues \m{\pm 1} on \m{{\cal H}_\pm}.
The restricted Unitary group is the subset of all unitaries satisfying a 
convergence condition:
\beq
	U_1({\cal H},\eps)=\{g|gg^{\dag}=g^{\dag}g=1;[\eps,g]\in {\cal I}_2\}.
\eeq
It  is vital for this definition to make sense that \m{{\cal I}_2} is an ideal;
 that is why the product of two elements still satisfies the convergence 
conditon.  (Any unitary operator is bounded). 
If  we split \m{g\in U_1({\cal H},\eps)} into
submatrices according to the splitting \m{g=\pmatrix{a&b\cr
				                     c&d\cr}}
where \m{a:{\cal H}_-\to {\cal H}_-} etc., the submatrices \m{b,c} are
H-S. The matrices \m{a,d} may not be invertible in general, but they
are Fredholm (see below)  since \m{g^{-1}} exists.

The restricted Grassmannian is the set of all self-adjoint operators of 
square one whose distance ( in the H-S sense) from \m{\eps} is finite.
\beq
	Gr_1({\cal H},\eps)=
\{\Phi|\Phi^\dag=\Phi;\Phi^2=1,\Phi-\eps\in {\cal I}_2\}.
\eeq
In the text we often use \m{M=\Phi-\eps} as the variable that describes a
 point in the Grassmannian. We showed that each such operator corresponds to 
 a subspace of \m{\cal H} ( the negative eigenspace of \m{\Phi}) 
  which is at a finite distance from the standard subspace \m{{\cal H}_-}.

The fact that \m{{\cal I}_2} is an ideal  ensures that  the action of
 the restricted unitary group on this  Grassmannian well-defined:
\beq
\Phi\mapsto g\Phi g^{\dag}, \quad M\to gMg^{\dag}+g[\eps, g^{\dag}].
\eeq

\section{Fredholm Index}

The material in this section is contained in Ref. \cite{presseg} to
which we refer for proofs and more precise statements.
A bounded  operator \m{A} is {\it Fredholm}  if it is invertible modulo a 
compact  operator; i.e., if there is a compact operator \m{K} such that 
\m{A+K} has an inverse. The set of Fredhom operators is a toplogical space, 
with the toplogy induced by the operator norm. It is disconnected, each 
connected component being labelled by an integer called the Fredholm index.

To understand the Fredholm index, consider the  kernel of an 
operator; i.e., the subspace of all \m{u} such that \m{Au=0}.
 Even in the finite
 dimensional case the kernel can change change discontinuosly under a small 
change in \m{A}.; for example we may cross an eigenvalue. In the 
finite dimensional  case, the dimensions of the kernels of \m{A} and 
\m{A^{\dag}} are the same. In the case of Fredholm operators on infinite 
dimensional Hilbert spaces, the difference
\beq
	{\rm index}\; (A)=\dim{\rm ker}\; A-\dim{\rm ker} A^{\dag}
\eeq
is always finite but need not vanish. It however does not change under
 continuos changes in \m{A}: it is constant in each connected
component of the  space of Freholm operators. In fact it is the only
such  function: any two Fredholm operators of the same index are
connected by a continuos curve.

Now \m{g\in U_1({\cal H},\eps)} is of course Fredholm of index zero:
it is invertible as is its adjoint. But the submatrices \m{a,d}( where
\m{a:{\cal H}_-\to {\cal H}_-} etc.) defined above are not invertible
in general. Yet they are invertible modulo some compact operators (
products such as \m{bb^{\dag}}), so they are Fredholm. Since \m{g} as
a whole invertible, they must have opposite Fredholm indexes. The index
of \m{a} (for example) is a topological invariant of \m{g}. The group
\m{U_1} is the union of connected components labelled by this index,
which can take any integer value. The connected component of the
identity, of course, has index zero. 

We can understand the renormalized dimension of a subspace (called
virtual rank in \cite{presseg}) in terms of the Fredhom index. Every
self-adjoint operator can be diagonalized; an operator \m{\Phi\in
Gr_1} can be brought to the standard form \m{\eps} by an element in
\m{U_1}: \m{\Phi=g\eps g^{\dag}}. The index of \m{a(g)} is then the
topological invariant associated to \m{\Phi}. 

Another point of view is in terms of the subspace \m{W} of \m{\cal H}
corresponding to \m{\Phi}. If \m{\Phi} is at a finite distance from
\m{\eps}, \m{W} will not differ `too much' from \m{{\cal H}_-}. More
precisely, the projection operator \m{\pi_+:W\to {\cal H}_+} will be
compact and \m{\pi_-:W\to {\cal H}_-} will be Fredholm. The index of
\m{\pi_-} measures the `difference in dimensions' between \m{W} and
\m{{\cal H}_-}; this is the renormalized dimension of \m{W}. We saw in
the text that this has the physical meaning of baryon number. 

It si crucial for all this that we allow only subspaces at a finite
distance from the standard one in \m{Gr_1}. If we had allowed for all 
subspaces, the Grassmannian would have been contractible!. The
convergence condition we must impose is required for the Poisson
brackets to make sense: the symplectic form of the Grassmannian would
not make sense otherwise. It is gratifying that as a consequence, we
get a topological invariant which has the physical meaning of baryon
number.

Although the symplectic form makes sense on all of the phase space
\m{Gr_1}, we should not expect the hamiltonian to make sense on all of
it. There should be some dense domain in which the hamiltonian does
make sense however. We leave such questions as challenges for the
analyst who is interested in solving problems of relevance to physics.

\chapter{  Quadrature  of Singular Integrals}

In this appendix we summarize some ideas on the numerical methods that are 
used in the chapter on solitons. Some originality is  needed  even 
in this part of the problem. 

\section{  Quadrature Formulas}

The equations we have are just too  hard to be solved analytically. We
have to resort to numerical methods. The basic idea is to convert the
integral equation into a matrix equation by allowing \m{p} to take
just a finite set of values:  we must find a way to
approximate the  integral by a finite sum. Then we will solve
the resulting nonlinear matrix equations by iteration.

Quadrature is the approximation of integrals by finite sums.
There are standard methods for quadrature, (method of
moments)  going back to
the days of Gauss. But we have to modify these methods since our
integrals are singular. The basic idea is well-known in the literature
on quadrature at least for the case of the Cauchy principal
value\cite{numint}.
 Our integrals are one step harder (Hadamard Finite Part) but
the idea is the same. See S. Chandrashekhar,'s classic book 
 \cite{radiative} for a clear discussion of numerical integration.

Let \m{\rho(x)} be a continuous positive   function on the close interval
\m{[a,b]}. We are interested in evaluating integrals such as 
\beq
	\int_a^b f(x)\rho(x)dx
\eeq
by numerical approximations. Here \m{f(x)} is some continuous function. 

Let \m{x_j, j=1,\cdots \nu} be a set of points in the interval
\m{[a,b]}. We expect a weighted sum  such as 
\beq
 \sum_j w_j f(x_j)
\eeq
to be good approximation for the integral, 
provided that (i)  the  number points \m{\nu} is large enough and (ii) the
points \m{x_j} are distributed  roughly uniformly.

 Every function can be approximated by a polynomial
within the interval; as the order \m{\nu-1} of the polynomials grows
the approximation gets better. We can thus approximate the integral of
\m{f(x)} by that of its polynomial approximation of order \m{\nu-1}.
The weights \m{w_j} are determined (once the points \m{x_j} are chosen)
by this  requirement:  the above formula should in particular  be exact for
polynomials of order \m{\nu-1}. 

This is the same as the condition
\beq
	\int_a^b x^k \rho(x) dx=\sum_j w_j x_j^k, 
\;{\rm for} \; k=0,1,\cdots k-1.
\eeq
The left hand side are the moments of the distribution \m{\rho(x)dx},
and are assumed to be known. Then the above set of linear equations
determine the weights in terms of \m{x_j}. 

If the function \m{\rho(x)dx} is not too rapidly varying, a simple
choice such as equally spaced  points \m{x_j=a+(b-a){j-1\over\nu}}
should give reasonable  approximation to the integral: certainly for
polynomials upto order \m{\nu-1} we will get the exact answer anyway.
 But it is
possible to do better by choosing the \m{\nu} points \m{x_j} cleverly, as
pointed out by Gauss:  we can
ensure that the answer is exact  for   polynomials of order
\m{2\nu-1}.  But we wont be using this idea: we will just use equal
spacing for the points, which turns out to be more convenient.
This because our  integrands involveterms such \m{\tilde\psi(p+r)}, so it is convenient if the sum two points \m{x_j+x_k} is also a point at which we evaluate the integrand.
 The
loss of precision in quadrature can be made up because the simplicity of equal spacing
allows us to choose a larger number of points.

\section{Singular Measures}

Now consider integral\cite{numint}  \m{{\cal FP}\int_0^b  f(x) {dx\over
x^2}}. The symbol \m{\cal FP} indicates as before  the
Hadamard  `finite part' of the integral.
For such singular integrals  we can also find a numerical
approximation as above. But the system of moments is no longer
positive. This is because the integation measure is no longer
positive: \m{{\cal FP}\int_0^b f(x){dx\over x^2}} can be negative even when
\m{f(x)} is positive.  The moments of the measure are given by 
\beq
	{\cal FP}\int_0^bx^k {dx\over x^2}={b^{k-1}\over k-1}\;{\rm
for}\; k\neq 1
\eeq
and 
\beq
	{\cal FP}\int_0^bx^k {dx\over x^2}=\log b\;{\rm
for}\; k=1.
\eeq
Note that the zeroth moment is negative. Also the second moment
violates scale invariance and is  negative if \m{b<1}.

Given a system of points \m{x_j} we can approximate the singular
integral by a sum 
\beq
	{\cal FP}\int_0^b f(x){dx\over x^2}=\sum_{j=1}^{\nu} w_j f(x_j)
\eeq
where the weights are determined as above by solving the system
\beq
	\mu_k=\sum_{j=1}^\nu w_j x_j^k\; {\rm for}\; k=0,\cdots \nu-1.
\eeq
The choice of equally spaced points gives good answers in many cases.

These methods are used in the text (towards the end of the second
chapter) to solve the integrals equations for the wavefunction of the baryon.

\backmatter

\end{document}